# Improving the crystallinity and texture of oblique-angle-deposited AlN thin films using reactive synchronized HiPIMS

Jyotish Patidar[1], Amit Sharma[2], Siarhei Zhuk[1], Giacomo Lorenzin[3], Claudia Cancellieri[3], Martin F. Sarott[4], Morgan Trassin[4], Kerstin Thorwarth[1], Johann Michler[2], Sebastian Siol[1,*]

[1]Laboratory for Surface Science and Coating Technologies, Empa – Swiss Federal Laboratories for Materials Science and Technology, Switzerland

[2]Laboratory for Mechanics of Materials and Nanostructures, Empa – Swiss Federal Laboratories for Materials Science and Technology, Switzerland

[3]Laboratory for Joining Technologies and Corrosion, Empa – Swiss Federal Laboratories for Materials Science and Technology, Switzerland

[4]Department of Materials, ETH Zürich, Switzerland

Corresponding author: sebastian.siol@empa.ch

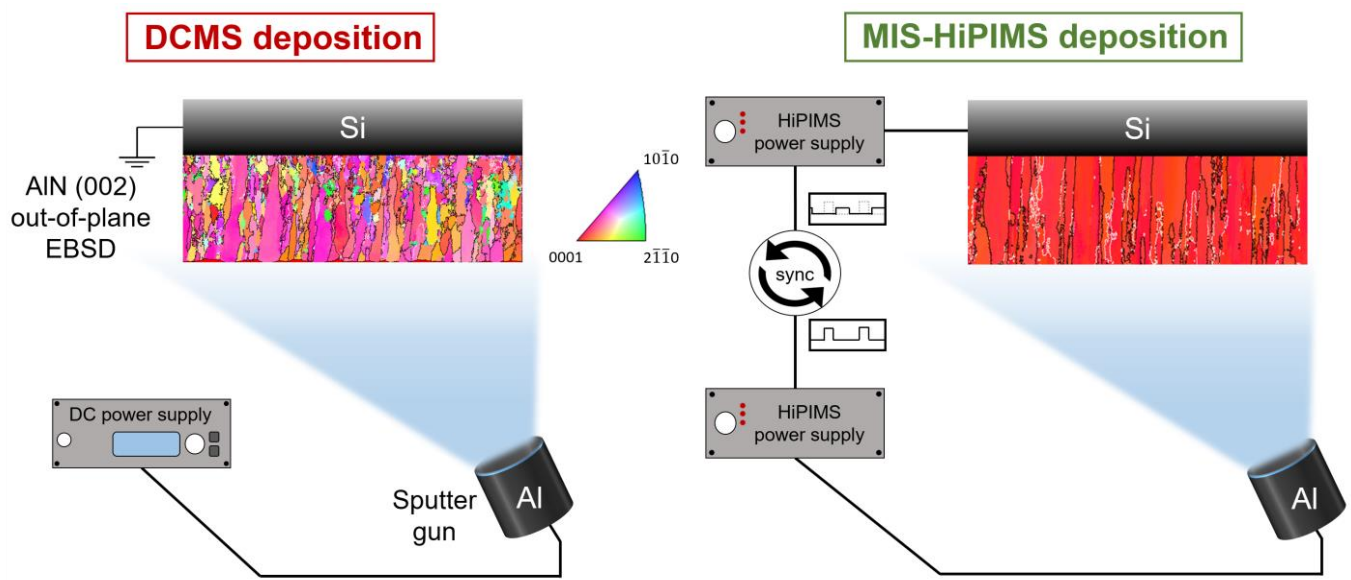

**Abstract:** Many technologies, such as surface-acoustic-wave (SAW) resonators, sensors, and piezoelectric MEMS require highly-oriented and textured functional thin films. The most common synthesis approaches use high deposition temperatures and sometimes epitaxial stabilization, whereas the best results are typically achieved for on-axis sputter geometries. In some scenarios, on-axis sputtering is not feasible, for instance, during co-deposition from multiple magnetrons or when coating structured substrates with high aspect ratios. During ionized physical vapor deposition (PVD), in contrast to conventional PVD, the film-forming species can be accelerated onto the growing film using substrate-bias potentials. This increases the ad-atom mobility but also deflects the trajectory of ions toward the substrate increasing the texture of the growing film. However, the increased amount of gas-ion incorporation in the films limits the feasibility of such synthesis approaches for the deposition of defect-sensitive functional thin films. In this work, we report on the oblique-angle deposition of highly textured, c-axis oriented AlN (0002) films, enabled by reactive metal-ion synchronized HiPIMS. The effect of critical deposition parameters, such as the magnetic configuration, ion kinetic energies as well as substrate biasing are investigated. AlN thin films deposited using direct current magnetron sputtering (DCMS) and conventional HiPIMS are discussed for comparison. The films deposited using HiPIMS show a more pronounced texture and orientation compared to DCMS films. We find that combining the HiPIMS depositions with a moderate substrate bias of only -30 V is sufficient to improve the crystalline quality and texture of the films significantly. To reduce process-gas incorporation, and the related formation of point defects, the negative substrate-bias potential is synchronized to only the Al-rich fraction of each HiPIMS pulse. This leads to reduced Ar-Ion incorporation and further improvement of the structural properties. In addition, to a pronounced out-of-plane texture, the films show uniform polarization of the grains making this synthesis route suitable for piezoelectric applications. While the compressive stress in the films is still comparatively high, the results already demonstrate, that metal-ion synchronized HiPIMS can yield promising results for the synthesis of functional thin films under oblique-angle deposition conditions - even with low substrate-bias potentials.

# 1. Introduction

The functional properties of thin-film materials are often governed by their structural properties. This makes the control of microstructure and texture an important challenge in the development of functional coatings. Physical vapor deposition (PVD) techniques, such as magnetron sputtering, enable the control of the microstructure over large ranges. Depending on the process parameters, different regions in the structure zone diagram varying from small porous crystallites to large columnar grains are accessible [1], [2]. Similarly, the texture of the films can be controlled by choosing appropriate growth conditions and choice of substrates [3]–[5]. For many applications, highly crystalline and textured films are preferred. In piezoelectric thin films, for instance, a pronounced and uniform out-of-plane texture of the films is necessary to achieve a high piezoelectric response [6], [7].

Reports have shown, that grain growth during sputtering is strongly influenced by the direction of the incident sputter flux [8]–[10]. During sputtering at oblique deposition angles, grains tend to preferably grow towards the deposition source. Consequently, for the growth of out-of-plane textured thin films, the preferred sputter geometry is along the substrate normal, i.e. "on-axis". In many deposition chamber designs, especially when co-sputtering from multiple magnetrons, on-axis deposition is not feasible. Moreover, some applications may also necessitate the deposition on structured surfaces. Here, the deposition angle changes with the surface topography, which makes the growth of uniformly textured films on these substrates a particularly challenging task [11]–[13]. Rotating the substrate during the deposition can eliminate the preferred growth axis, but the continuously varying deposition angle can lead to a reduction in overall texture and crystallinity, when compared with on-axis deposited films.

Ionized PVD (IPVD) methods, particularly high-power impulse magnetron sputtering (HiPIMS) have gained interest in recent years for several applications [14]–[16]. IPVD techniques are characterized by their high ionization rates compared to conventional sputtering methods, which open up exciting opportunities for process design [16]–[18]. HiPIMS is a type of IPVD method, in which the power to the sputtering target is applied in the form of highly energetic, but short pulses in the microsecond range. As a consequence, it is possible to reach much higher power densities without increasing the thermal load on the target or substrate. In addition, the high plasma densities result in higher ionization fractions, which can even exceed 90% [19], [20]. The abundant ions in the plasma can be accelerated to the growing film with the application of substrate bias potentials, thus achieving higher ad-atom mobility without requiring high growth temperatures, resulting in denser and often more crystalline coatings at lower temperatures [21], [22]. This is particularly interesting for the deposition on temperature-sensitive substrates [23]–[25]. In addition, the trajectory of the ions can be altered by the bias potentials facilitating the growth of uniform and textured films during oblique-angle deposition or on substrates with high aspect ratios [26], [27]. **Figure 1** schematically shows the oblique-angle-deposition of films in direct current magnetron sputtering (DCMS) and HiPIMS with the addition of a negative substrate bias potential. Whereas the DCMS-deposited film tends to grow in the direction of the incident sputter flux, a highly oriented film is deposited with the aid of directed ion-irradiation using a negative substrate bias potential in HiPIMS. The tilting of grains in oblique-angle depositions has been previously reported in multiple works, whereas the tilt can be approximated using the empirical cosine and tangent rules [28], [29].

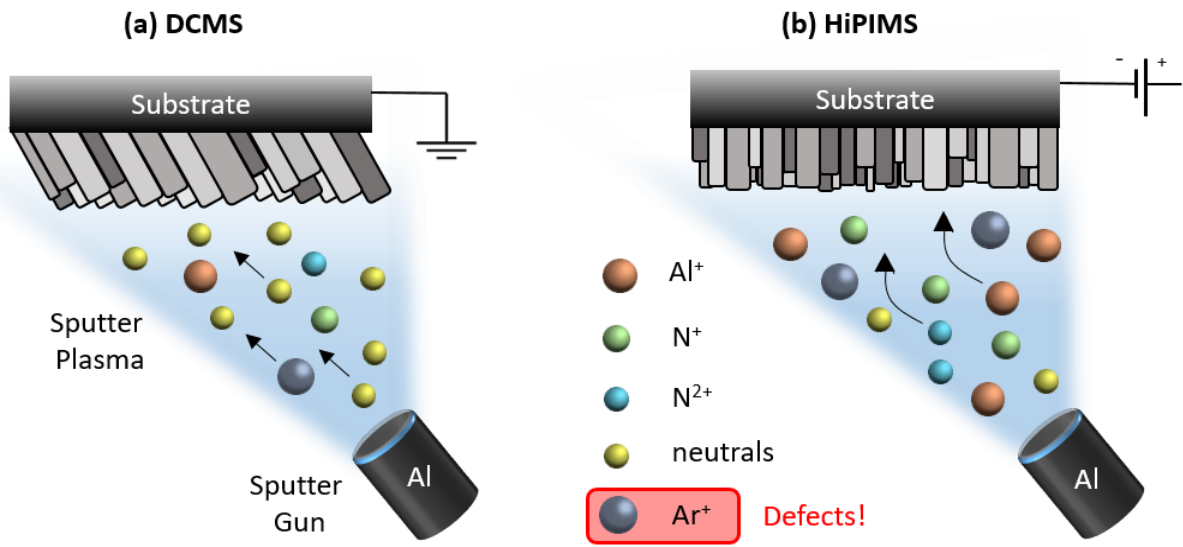

**Figure 1.** Schematic representing growth of AlN films in (a) DCMS and (b) HiPIMS methods. The grains are oriented in the direction of the incident sputter flux in DCMS, while a columnar growth normal to the substrate plane can be observed in HiPIMS-deposited films using negative substrate-bias potentials due to directed ion-irradiation along the substrate normal. At the same time, acceleration and implantation of Ar$^+$ ions can lead to process gas incorporation in the film leading to the generation of defects.

To date, HiPIMS is mainly used for the deposition of hard coatings and metallic thin films [30]. Especially for the deposition of wear- and temperature-resistant ceramics, such as TiAlN, TiAlSiN, CrAlTiN, and CrWN, this method is well established [31]–[34]. During HiPIMS of these types of coatings substrate bias potentials exceeding -100 V are commonly used. The resulting high adatom mobility, as well as the subplantation of ions with high kinetic energy promote the formation of dense films and often result in mechanically stronger coatings compared to those prepared with conventional sputtering techniques [35]. However, one of the downsides of this approach is that the negative substrate bias also accelerates the process gas ions. These ions, which are abundant in the plasma, can be implanted in interstitial sites causing high amounts of compressive stress in the films [2], [36]. This, in turn, can lead to the disorientation of grains and limit the texture of the film. The presence of such defects in the deposited films is less consequential for the mechanical properties, compared to the optoelectronic properties. By breaking the periodicity of the crystal, these defects can result in the formation of barriers for charge transport and thus reduced charge carrier mobility [37]. In addition, these defects can also act as recombination sites, reducing carrier lifetimes and breakdown potentials [38], [39].

With recent developments and the introduction of novel synthesis approaches like metal-ion-synchronized HiPIMS (MIS-HiPIMS), it is now possible to selectively increase the kinetic energy of specific ions, while at the same time minimizing process gas incorporation in the films [40], [41]. During HiPIMS, different sputtered species can arrive at the substrate at different times following each pulse. Gas-rarefaction effects, as well as differences in masses result in different time-of-flight for different ionic species. This causes a phase shift between the gas- and the metal-rich parts of the sputter flux incident at the substrate, i.e., in each pulse the gas-ions arrive before the metal-ions [42]–[45]. MIS-HiPIMS utilizes this effect to selectively accelerate specific species. If the arrival time of the different ions at the substrate is known, the bias potential can be pulsed accordingly to attract the metal-ion-rich part of the plasma, essentially providing the advantages of energetic ion bombardment during the film's growth and at the same time reducing the gas incorporation and thus defects in the films. Several recent research works have shown improvements in crystallinity and texture with reduction of the gas-ion incorporation in the films with the application of synchronized HiPIMS [46]–[49]. This raises the question if MIS-HiPIMS approaches can be used for a wider variety of coatings, such as electroceramics or even semiconducting coatings.

In this work, we investigate the advantages of synchronized HiPIMS over conventional sputtering approaches for the deposition of AlN thin films in an oblique-angle deposition geometry. Since AlN is commonly used in piezoelectric applications the crystallographic texture is one of the most critical properties when synthesizing AlN thin films.

Aluminum nitride in wurtzite structure (space group 186, $P6_3mc$) is one of the most widely employed materials for piezoelectric applications, mainly because of its unique properties such as linear frequency response, piezoelectric stability at higher temperatures, wide band gap, and compatibility with CMOS technology [50], [51]. A variety of deposition techniques has been used to produce textured AlN thin films ranging from direct current magnetron sputtering (DCMS), HiPIMS, chemical vapor deposition, and atomic layer deposition [13], [52]–[55]. In industrial settings, AlN is commonly synthesized using DCMS in an on-axis geometry using large targets and relatively low working distances on substrates of low surface roughness to achieve high deposition rates and pronounced out-of-plane texture, and low compressive stress.

In this study, we synthesize AlN film in oblique-angle geometry in a confocal sputter-up geometry as it is found in many R&D sputter chambers worldwide. Moreover, with this deposition setup, the results and their implications can also be applied to the deposition of films on structured substrates. Different approaches for the reactive sputter deposition are used including DCMS and synchronized HiPIMS. The process optimization and development are discussed in detail along with an in-depth characterization of the resulting film properties. The films deposited using HiPIMS show an enhanced texture and orientation compared to the DCMS films deposited with similar conditions. The effect of low energy ion-irradiation is investigated by working with small substrate-bias voltages to reduce the amount of defects, which are crucial for piezoelectric applications. Furthermore, the application of synchronized substrate biasing is investigated. XRD measurements show a strongly pronounced c-axis orientation, whereas piezoresponse force microscopy confirms a uniform polarization of the grains, which highlights the promise of MIS-HiPIMS for the deposition of piezoelectric thin films.

## 2. Methods

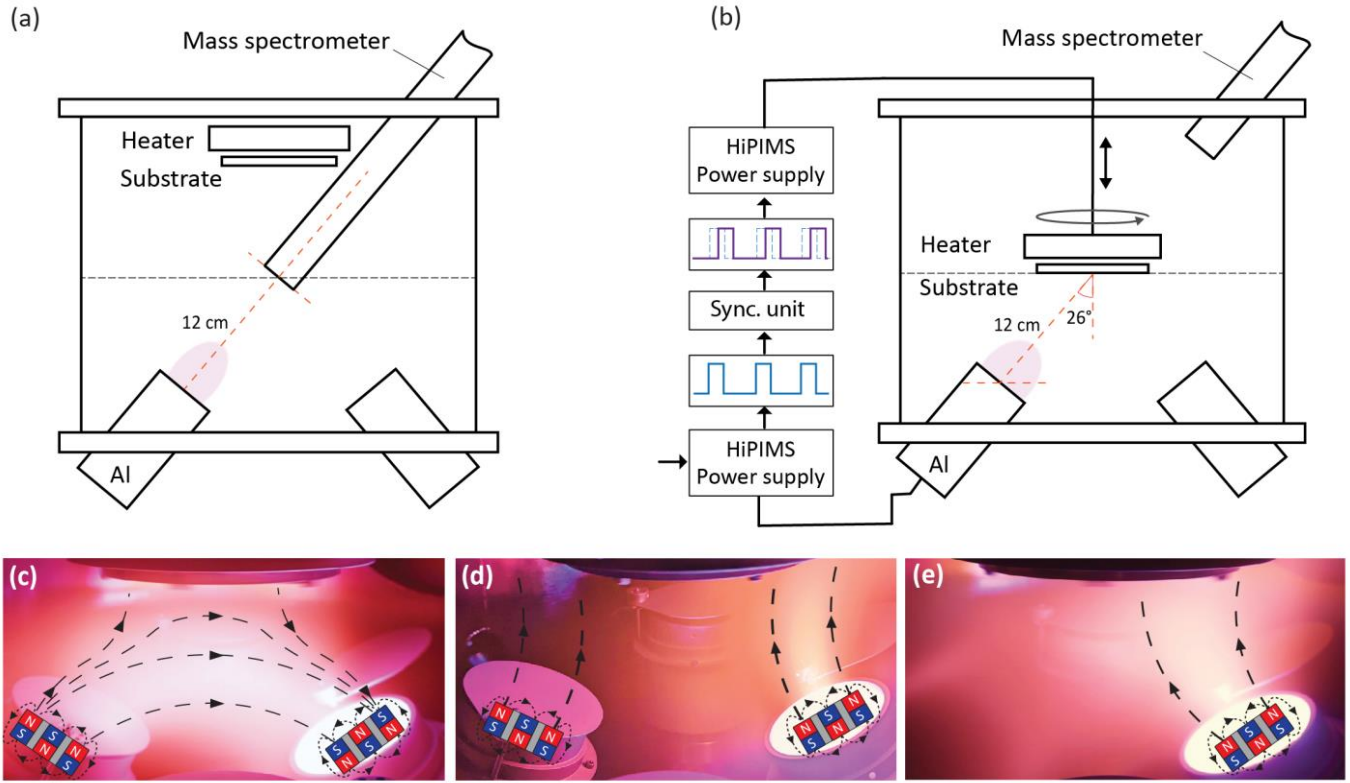

**Figure 2:** Schematic of the chamber used in this study. (a) The chamber is equipped with a mass spectrometer, which is moved to the working distance to perform time- and energy-resolved measurements of the incident ions. (b) The spectrometer is retracted and the substrate holder with heater is moved to do the deposition at the same working distance. The distribution of the sputter plasma can be changed by changing the magnetic configuration in the chamber. **Figures (c)** and **(d, e)** represent the closed-field plasma and open-field plasma, respectively. The plasma distribution can be further tuned by moving the magnetrons towards or away from each other **(d, e)**.

AlN thin films were deposited using DCMS and HiPIMS in a commercial, custom-built sputter chamber (AJA International, ATC-1800), as shown in **Figure 2**. The system has four sputter guns arranged in a confocal sputter-up geometry. In deposition systems with multiple magnetrons, the magnetic fields of different sputter guns can interact. By modifying the magnet configuration of each gun the magnetic field distribution inside the chamber can be altered [56]–[59]. In this study, we investigate two common magnetic configurations, namely, open and closed, represented in **Figure 2 (c-e)**. The magnetic configuration of the active magnetron was changed from a closed field (CF) to an open field (OF) configuration with the opposing counterpart, resulting in different plasma distributions in the chamber. In a CF configuration, the plasma is confined along the "closed" magnetic field lines of two magnetrons with opposite magnetic polarization. In an OF configuration, the magnetic configuration of the magnetrons is identical, so that the magnetic field lines and consequently the plasma is guided toward the substrate.

The substrate holder is either grounded or negatively biased based on the different approaches followed in this work. Heating is achieved with 5 halogen lamps, resulting in homogenous temperature distribution on the substrate. The DCMS depositions were carried out using a 750W DC power supply (AJA International, DCXS 750). HiPIMS depositions were carried out with Ionautics pulsing units and power supplies (Ionautics, HiPSTER 1 bipolar). The pulsed substrate bias was applied using the same.

The depositions were carried out on p-type (001) Si wafers from a single unbalanced magnetron equipped with a 2 inch Al target (HMW Hauner, purity: 99.999 at.%) with a sputter angle of 26° (with respect to substrate normal) and a working distance of 12 cm. The substrates were ultrasonically cleaned in acetone and ethanol before the deposition. A base pressure of < 1×10$^{-8}$ mbar was achieved before the deposition to ensure minimal contamination from the chamber during the deposition. The films were deposited at a time-averaged power of 100 W and 3 μbar working pressure. The chamber is equipped with a set of diverter valves (supporting information Fig. S1) through which the gases can be routed to specific sputter guns. Based on the hysteresis studies (in supporting information Fig. S2), the working gas, Ar, was routed away from the active sputter gun and reactive gas, $N_2$, to the sputter gun. The substrate temperature was maintained at 280 °C throughout the deposition. Substrate rotation of approx. 15 rpm was used to ensure uniform deposition. The deposition time was varied for DCMS and HiPIMS samples to facilitate similar thicknesses of the films. Magnetrons, which were not used were tilted towards the chamber walls to minimize the influence of their magnetic field. All deposition parameters are summarized in **Table 1**.

**Table 1.** Deposition parameters for DCMS- and HiPIMS-deposited AlN films in this study.

| Parameter | DCMS | HiPIMS |
|---|---|---|
| Target-substrate distance (cm) | 12 | 12 |
| Sputter angle (°) | 26 | 26 |
| Avg. power density (W/cm²) | ~ 5 | ~ 5 |
| Peak current density (A/cm²) | 0.015 | 0.5 |
| Working pressure (μbar) | 3 | 3 |
| Ar/N₂ flow rate (sccm) | 20/10 | 20/10 |
| Substrate temperature (°C) | 280 | 280 |
| Pulse width (μs) | - | 10 |
| Frequency (kHz) | - | 7.5 kHz |
| Substrate bias | ground | ground / -30 V DC / -30 V pulse for 40 μs (MIS-HiPIMS) |
| Deposition rate (nm/min) | 1.5 | 0.8 |

The ion energy distribution functions (IEDF) of the aluminum, nitrogen, and argon positive ions were measured using an energy and time-resolved mass spectrometer (EQP-300 Hiden Analytical) (see Figure 2b). The electrically grounded orifice (30 μm in diameter) of the instrument was placed at the working distance while facing the sputter gun. For the time-resolved measurements, the triggering signal was provided from the pulse generator of the HiPIMS power supply and used to synchronize the measurements with the HiPIMS pulse. Ion energy distribution functions (IEDF) of 0-50 eV were measured with a delay time of 10 μs/step. The gate width (i.e. the time for which the detector collects the ions at every step) was set to 10 μs, consistent with the chosen step size. The profiling was done by measuring each ion species separately. For the synchronization of the substrate bias potentials, the HiPIMS power supplies for the target and substrate were connected to a synchronization unit (Ionautics). The trigger from the HiPIMS pulse on the target was used to synchronize the substrate bias pulse with the metal-rich phase of the HiPIMS plasma. To avoid saturation of the detector, a less abundant $^{36}$Ar isotope was used for these measurements.

The transit times of ions in the mass spectrometer was obtained using the equations suggested in literature and were calculated to be 108 µs for $Ar^+$, 93 µs for $Al^+$, 67 µs for $N^+$ and 95 µs for $N_2^+$ [60]. The substrate potential was actively regulated to 0 V between the substrate bias pulses.

X-ray diffraction (XRD) analysis of the films was performed using a Bruker D8 in Bragg Brentano geometry and Cu $k_\alpha$ radiation. Pole figures are acquired around the AlN {0002} and {10$\bar{1}$3} families of planes to check the in-plane and out-of-plane texture. The pole figure scans were performed for psi angles ranging from 0° to 80° and phi from 0° to 360° with a step size of 3°. Stress analysis is performed using the Crystallite Group Method (CGM), suitable for highly textured layers [61]. The thickness of the films was measured using a Dektak profilometer. The Ar-ion incorporation was estimated using energy-dispersive X-ray spectroscopy (EDS) measurements performed in a Hitachi S3700 system equipped with an EDAX octane pro detector. An acceleration voltage of 5 kV was used resulting in a probing depth of about 500 nm.

The cross-sections for transmission electron microscopy (TEM) were prepared by focused ion beam milling (Tescan, Lyra3). The STEM high-angle annular dark field (HAADF) imaging, TEM Bright field/dark field imaging and selected-area diffraction (SAED) studies were carried out using a Themis 200 G3 aberration (probe) corrected TEM (Thermo Fischer) operating at 200 kV. Orientation imaging along the cross-section of the films was performed with scanning precession electron diffraction (SPED) technique in the TEM-based orientation imaging microscopic analysis using NanoMegas (Digital/ASTAR) having a resolution of the order of a nanometer [62]. A step size of 7 nm and a precession of 0.8° was used for all the measurements. The roughness and morphology of the films were measured using Bruker nanoscope AFM using ScanAsyst mode with silicon cantilever tips. $1 \times 1$ µm$^2$ AFM images with 512 × 512 pixel resolution were obtained with Nanoscope software and later processed in Gwyddion. Piezoresponse force microscopy measurements were performed using a Bruker Multimode 8 atomic force microscope, equipped with Pt-coated Si tips (k=5.4 N/m, MicroMasch). The piezoresponse was measured at a drive frequency of 10 kHz and a drive amplitude $V_{AC}$ of 5 V.

The oxygen levels in the films were measured via depth profiling using a PHI-Quantera X-ray photoelectron spectroscopy (XPS). The XPS system is equipped with Al kα radiation. Charge neutralization was used throughout the measurement using a dual-beam neutralizer. The films were transferred to the XPS chamber from the deposition chamber using a custom-built UHV-transfer cart, where the pressure was maintained below $10^{-8}$ mbar throughout the transfer process. The XPS spectra and profiling were recorded on samples that were transferred in UHV conditions and then later exposed to the atmosphere for 5 minutes, to measure the oxidation in ambient conditions.

# 3. Results and Discussion

### 3.1 Characterization of the HiPIMS discharge and synchronization of the substrate bias potential

The plasma parameters and discharge properties during the deposition play an important role in determining the growth of the film and subsequently its properties [63]–[65]. The plasma in the deposition chamber was first characterized using the mass spectrometer with a grounded aperture to perform measurements of the ion energy distribution function (IEDF). The IEDF of the predominant incoming ionic species present in the plasma for DCMS and HiPIMS techniques are shown in **Figure 3**. While not quantitative, the difference in ion count rate is striking with HiPIMS ion count rates exceeding those of the DCMS discharge by two orders of magnitude. The median kinetic energy of Al$^+$ ions is increased from approximately 6.6 eV to 10.7 eV, for DMCS and HiPIMS processes, respectively. Measurements of the N$^+$-ions also follow the same trend. The flux of doubly-ionized ions was also measured, however, the amount was much lower than singly-ionized species and thus their effect on film growth is expected to be minimal. In HiPIMS, we can see different populations of ionic species for Al, Ar, and N ions in different energy ranges. The low energy peaks correspond to the background signal. The major peak and the following tail for the Al and N correspond mainly to the ions generated during the pulsed HiPIMS discharge from the nitrided surface of the target. The contribution of metal ions to the total ion flux during HiPIMS mainly depends on the pulse parameters, target state, and target current density [66], [67]. The supply of reactive gas directly to the vicinity of the target leads to high dissociation and ionization of nitrogen molecules. It is also interesting to note, that the amount of dissociated nitrogen ions is considerably higher in HiPIMS. This is consistent with observations by other groups [68], [69].

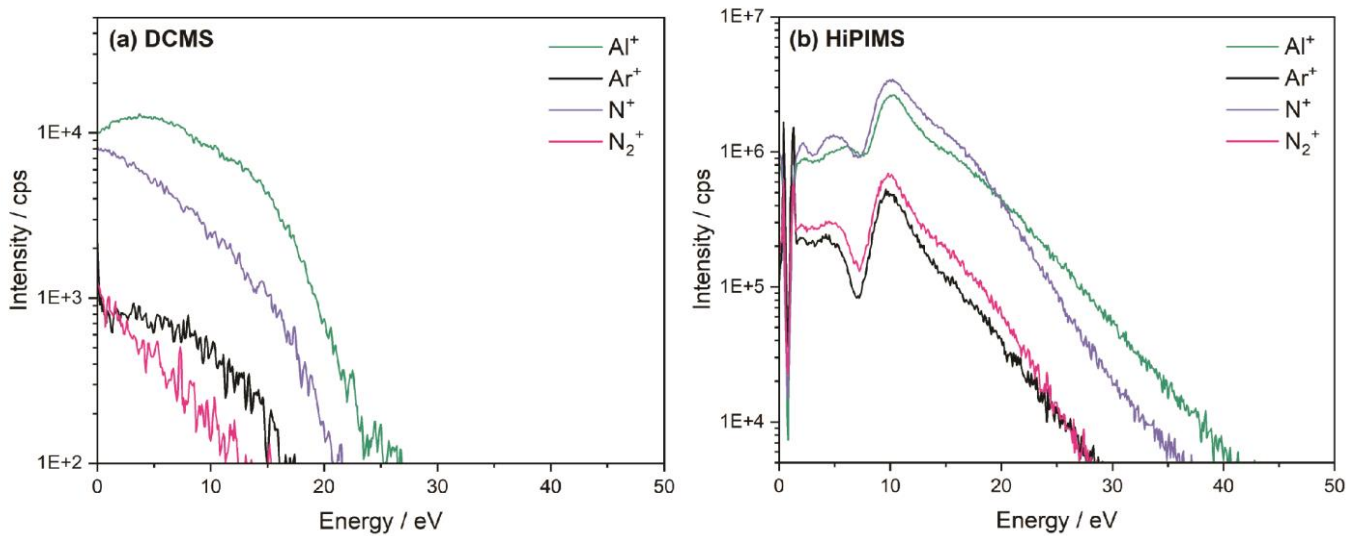

**Figure 3:** Time-averaged ion energy distribution functions (IEDFs) measured for (a) DCMS and (b) HiPIMS plasma. As expected, the ion flux and kinetic energies measured in HiPIMS are significantly higher compared to DCMS.

In order to set up the MIS-HiPIMS process, we performed measurements of the IEDF of different ions at different delay times relative to the applied HiPIMS pulse. Integrating the IEDF for each delay time, provides the ion flux as a function of time and consequently the ion's time-of-flight (see Section S3 of supporting information). **Figure 4** shows the I-U curves for the applied HiPIMS pulse on the Al target as well as the integrated ion fluxes of Ar$^+$ and Al$^+$ incident on the mass spectrometer. Clearly, the Ar$^+$-ions tend to arrive at the substrate earlier than the Al$^+$-ions.

Based on this information, the substrate pulse is tailored to preferentially accelerate the metal ions toward the substrate. It is important to note here that the substrate bias is actively regulated to 0 V between the metal-ion-rich pulses reducing an otherwise pronounced self-bias potential. The gas-ions incident at this time of the pulse have energies below the lattice displacement threshold, and thus the structure of the film should primarily influenced by the ion irradiation and momentum transfer from the metal-rich part of the HiPIMS pulse [70].

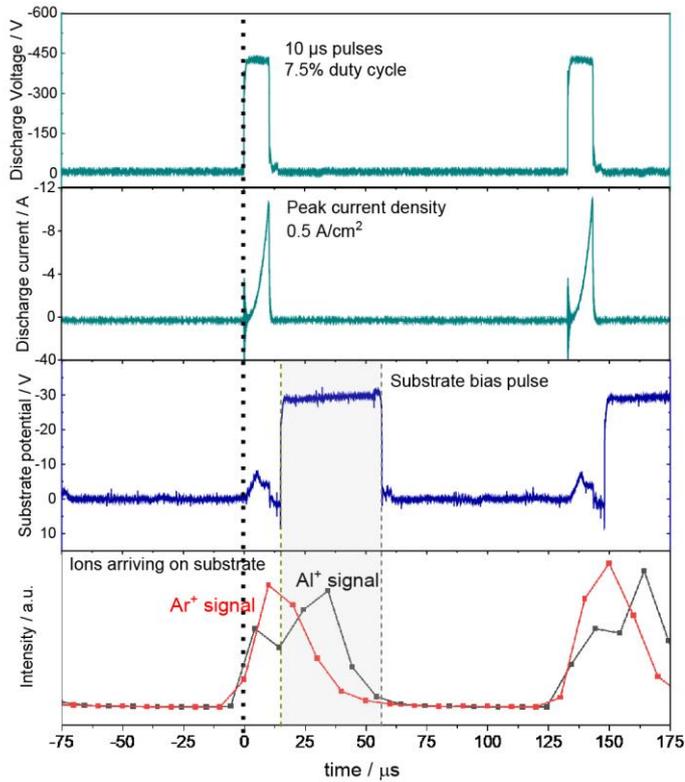

**Figure 4.** Synchronization of substrate pulse with HiPIMS pulse. The substrate bias pulse is tailored to attract the metal-ions selectively for energetic metal-ion bombardment during the film growth, simultaneously avoiding gas-ion incorporation in the films.

## 3.2 Structural characterization of the films

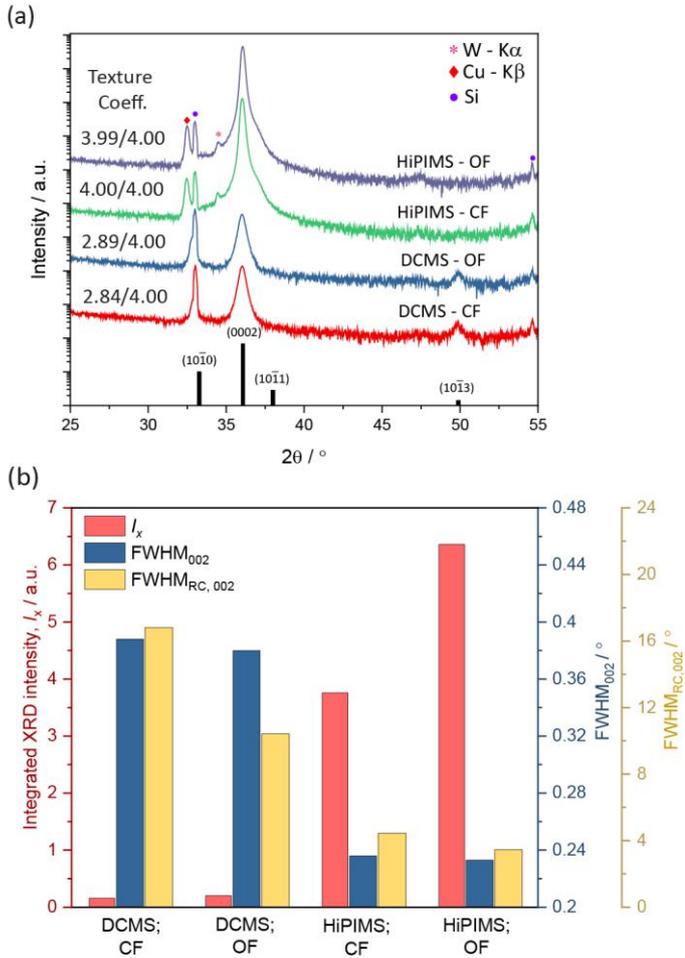

**Figure 5:** (a) XRD patterns of films deposited using DCMS and HiPIMS technique with different magnetic configurations in the chamber (b) Analysis of crystalline quality of the films by comparison of integrated intensity $I_x$ and FWHM of the rocking curves for different magnetic configurations. The textural quality of the film is seen to improve with open-field depositions.

In plasma-based physical vapor deposition processes, phase formation and nucleation can be greatly influenced by changes in the magnetic field distribution [56], [71]–[75]. In an initial assessment of the deposition-chamber setup, we evaluated the influence of the magnetic field distribution on the structural properties of the AlN thin films. XRD was performed on AlN thin films deposited using DCMS/HiPIMS in both open-field (OF) and closed-field (CF) configurations. The respective XRD patterns are shown in **Figure 5a**. All films show a pronounced out-of-plane c-axis orientation. For a rough estimation of the crystalline quality of the films, we used the integrated XRD intensity $I_x$, which is defined as the integrated area of the (0002) diffraction peak normalized by the thickness of the film. The variation of $I_x$ for samples deposited with different techniques and magnetic configurations is plotted in **Figure 5b**. The crystallinity and thus $I_x$ increase by using an OF configuration during the deposition, irrespective of the deposition technique. Additionally, the crystalline quality of films deposited with HiPIMS is considerably higher than the DCMS films, owing to the higher ion kinetic energies in the case of HiPIMS. In OF configuration a larger amount of ions are accelerated toward the film. This increases the fraction of ions in the film-forming sputter flux, but also leads to additional plasma heating. Both effects have been shown to improve the adatom mobility and consequently facilitate crystallization during film growth [76]–[78].

This is also reflected in an improved in-plane crystal coherency and mosaicity of the grains in the films which is evaluated by performing rocking curve measurements on the (0002) reflection. The rocking curves can be found in the Section S4 of supporting information. The FWHM of the rocking curve is often used as the relative measure of c-axis orientation. Here, a lower FWHM indicates fewer misaligned grains and a more coherent in-plane crystal arrangement.

Low energy ion-irradiation during the film growth results in higher adatom mobility, allowing the atoms to move to the lattice sites closest to equilibrium, i.e., a close-packed (0002) basal plane with the lowest surface energy [79]. This, in turn, leads to a more pronounced out-of-plane texture and a fiber texture in-plane. The crystalline quality and texture improvements with OF configuration are more pronounced in HiPIMS as compared to DCMS-deposited samples owing to the higher ionization and ion irradiation rate. The ion bombardment can be further tuned to increase the energy of impinging ions by changing the substrate bias potentials, which is discussed in later sections. Due to the clear improvements in crystallinity and texture, the following experiments reported in this work are performed in OF magnetic configuration unless specifically stated otherwise.

For further investigations, we compared four representative samples deposited with different approaches which are: AlN film deposited with DCMS and a grounded substrate, HiPIMS with a grounded substrate, HiPIMS with a continuous bias potential of -30 V on the substrate as well as HiPIMS with the synchronized pulsed substrate bias potential of -30 V for a duration of 40 µs. The bias potentials are intentionally chosen conservatively to avoid Ar-ion incorporation in the films. All samples are deposited with open field configuration and exhibit less than 1 at.% oxygen contamination, as measured by XPS without exposing the films to the atmosphere. The XRD patterns, rocking curves, $I_x$, as well as the FWHM of the (0002) diffraction peak and rocking curve are plotted in **Figure 6a-c**. The energy of the impinging species during the film growth is further increased by the application of substrate bias, resulting in better crystalline properties. A clear difference in the peak intensity is observed by changing the deposition method from DCMS to HiPIMS. Texture coefficient is also increasing remarkably in the HiPIMS films. All HiPIMS films are nearly perfectly textured. Since 4 diffraction peaks are considered in the analysis, a texture coefficient of 4 would denote a completely textured film [80]. A shift of the (0002) diffraction peak towards lower 2θ values with increasing substrate bias indicates an increase in compressive stress in the films. This effect is most pronounced for the continuous substrate bias potential. A more detailed discussion of the stress state is given in the following section.

The rocking curve on the (0002) reflection was measured to estimate the mosaicity of samples (see **Fig. 6b**). The FWHM of the rocking curve is seen to reduce as we move from DCMS to HiPIMS and further to continuous bias and synchronized bias, indicating an improvement in the texture of the films. In addition, for a more comprehensive assessment of the texture, XRD pole figures were recorded for the (0002) and (10$\bar{1}$3) reflections (**Fig. 6d**). The pole figures confirm the latter results with the MIS-HiPIMS film being the most strongly oriented in (0002). The films exhibit a fiber texture i.e. the films are randomly oriented in-plane with a preferential orientation of grains normal to the plane of the substrate. The narrower intensity distribution in MIS-HiPIMS film indicates increased in-plane crystal coherency.

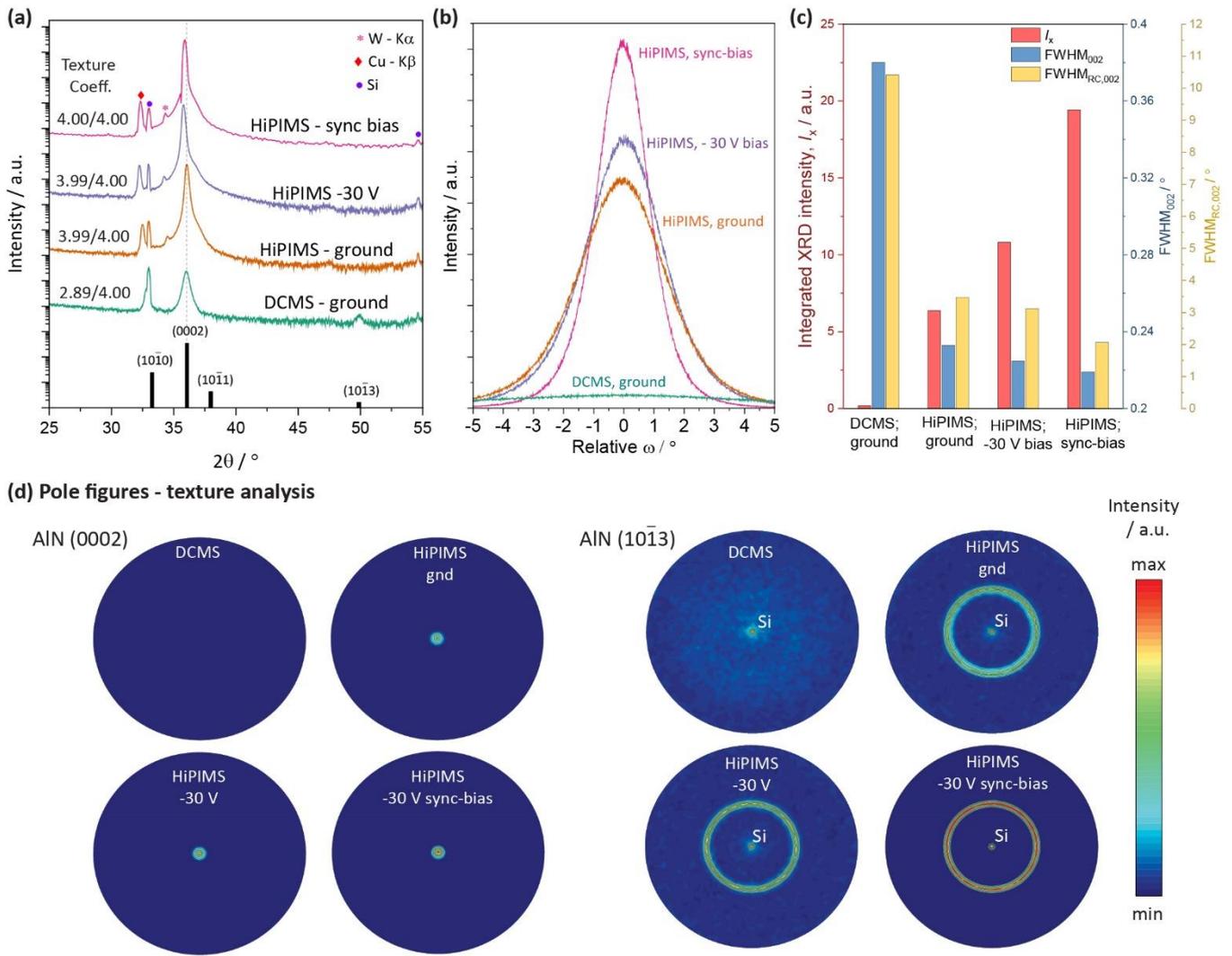

**Figure 6:** Structural analysis of the representative films. (a) XRD patterns of films deposited with different approaches; (b) Normalized rocking curves; (c) Variation of integrated intensity $I_x$, FWHM of (0002) diffraction peak, and Ar incorporation in the deposited films. The crystalline quality is markedly improved with HiPIMS and furthermore with the application of biasing, particularly synchronized biasing. (d) XRD pole figures of the (0002) and (10$\bar{1}$3) reflections. The pole figures suggest increased out-of-plane texture in (0002) direction and random in-plane texture for the HiPIMS deposited films.

### 3.3  Stress analysis of the films

The thin film stress state was evaluated using the Crystallite Group Method (CGM) [81], [82]. Since the films in this study are textured, the XRD reflections of planes apart from (0002) were measured by tilting the sample to a specific psi angle, Ψ, corresponding to the difference in angles between the (0002) and respective plane. The in plane angle phi is not critical as the samples show a fiber textured in plane (see pole figures). The XRD patterns for the four representative samples can be found in section S5 of the supporting information. The stress is further calculated using the elastic coefficients (compliance constants $S_{ij}$) and equations suggested in the literature [83]–[86]. The total calculated values of stress are plotted in **Figure 7** along with the Ar incorporation in the film from EDS analysis. Here negative values of stress correspond to compressive stress in the system. All films exhibit large amounts of compressive stress, where the highest stress values were measured for the HiPIMS depositions using substrate biasing.

Several factors contribute to stress in thin films. The most common factors are thermal, epiaxial as well as intrinsic stresses. All films were deposited using a nominal substrate temperature of 280 °C. During sputtering the actual surface temperature on the substrate can be significantly higher (e.g. due to plasma heating or condensation heat transfer). Assuming a maximum temperature of 600 °C as well as the difference in thermal expansion coefficients for Si and AlN of $\Delta\alpha\approx1.6^{-6}$/K, the thermal stress of the films would be on the order of a few hundred MPa and therefore cannot explain the observed stress evolution. The effect of epitaxial stress is potentially larger. Valcheva et al. reported the epitaxial growth of AlN (0001) on Si (001) using reactive sputtering. Depending on the respective domain orientation of AlN a significant lattice mismatch to the underlying substrate is observed. The formation of misfit dislocations can relieve this strain to values of less than 1%. However, their results indicated that the strain is predominantly tensile [87]. This stands in contrast with our observations that the films with the most pronounced in-plane orientation exhibit the highest compressive stress values. While the exact origin of the high compressive stress in the films is hard to pinpoint, the main contributing factor for the increased compressive stress in the films appears to be intrinsic. It is noteworthy, that while the lower crystallinity as well as open columnar structure of the DCMS films provide opportunities for stress relaxation, the HiPIMS films without substrate bias show lower compressive stress, while at the same time exhibiting higher crystallinity as well as a compact microstructure as evidenced in later sections.

The most striking increase in compressive stress occurs with the application of a substrate bias and therefore is likely related to the energetic ion bombardment during growth. Upon application of a substrate bias potential, the kinetic energy of the incident ions is increased by the plasma-potential drop at the substrate. Based on the median kinetic energies during HiPIMS on the order of 10 eV (see Figure 3) this would produce a significant amount of ions with kinetic energy over 40 eV, i.e. above the lattice displacement threshold [88]. This can lead to ion-implantation and structural defects in the growing films as reported by several groups [30], [89]. Due to the improved crystalline quality of the HiPIMS films, the increase in stress is more likely due to ion-implantation. It was observed that the compressive stress in the films strongly correlates with the Ar incorporation. In contrast to the film-forming species, the energetic $Ar^+$ ions from the plasma are incorporated at interstitial sites in the AlN lattice, so that even small amounts of impurities can generate a significant amount of stress in the films [49], [90].

Through appropriate substrate-bias synchronization, this detrimental effect can be mitigated. While the Ar content and compressive stress in the MIS-HiPIMS film are lower than for the constant bias deposition, there is still room for further improvements by optimizing the substrate bias pulse pattern.

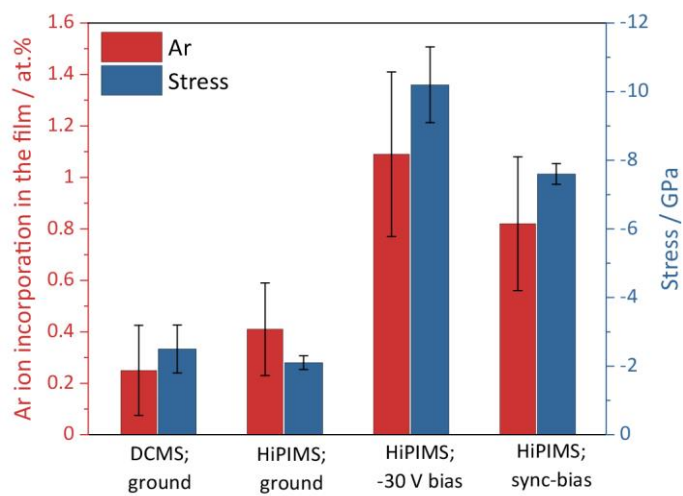

**Figure 7:** Evaluation of the in-plane stress in the films deposited with different approaches. The compressive stress in the films is strongly correlated with the Ar-ion incorporation in the films. The amount of Ar incorporation in films decreases with the application of the synchronized bias.

### 3.4 Microstructure analysis of the films

Transmission electron microscopy (TEM) cross-sections of the films were analyzed on focused ion beam (FIB) lamellae from the four representative samples (see **Fig. 8 and Section S6 of the Supporting Information)**. Bright-field TEM images are shown along with the corresponding orientation maps obtained via SPED. All AlN films exhibit a columnar structure. Grain size and texture are increased for the HiPIMS films as compared to the DCMS film, which is also visible in the SPED analysis of the respective films. The reduced grain size is in line with the lower ad-atom mobility during the DCMS process. HiPIMS films, especially the films deposited with continuous and pulsed biasing show a compact columnar growth. Upon adding the substrate bias the grain size is further increased with grains extending over the full thickness of the film.

A substantial difference in the texture of the films is evident from in-plane as well as out-of-plane SPED analysis. The DCMS film exhibits generally small randomly oriented grains, while a highly oriented (0002) film can be seen for HiPIMS with synchronized biasing. While the pole figures of the AlN $(10\bar{1}3)$ reflection shows no pronounced in-plane orientation, the more sensitive SPED measurement reveals a weak preferential in-plane orientation in $(2\bar{1}\bar{1}0)$ and $(10\bar{1}0)$ directions. This is well in line with the epitaxial relationship between Si (001) and AlN (0001) [87], [91]. To further analyze the texture of the films selected area electron diffraction (SAED) was performed. As expected, the DCMS film shows a randomly oriented polycrystalline structure while the HiPIMS film with synchronized biasing shows a bi-axial texture along the $(2\bar{1}\bar{1}0)$ and $(10\bar{1}0)$ zone axes. It is noteworthy, that as we move from DCMS to HiPIMS and to additional substrate biasing, the kinetic energy of the film forming ions is increasing while the nominal substrate temperature is kept constant. The more pronounced columnar growth with the increased energetic bombardment is well in line with the microstructural evolution outlined in the often referenced modified structure zone diagram [2].

**(a) Transmission electron microscopy (TEM) and scanning precession electron diffraction (SPED) imaging**

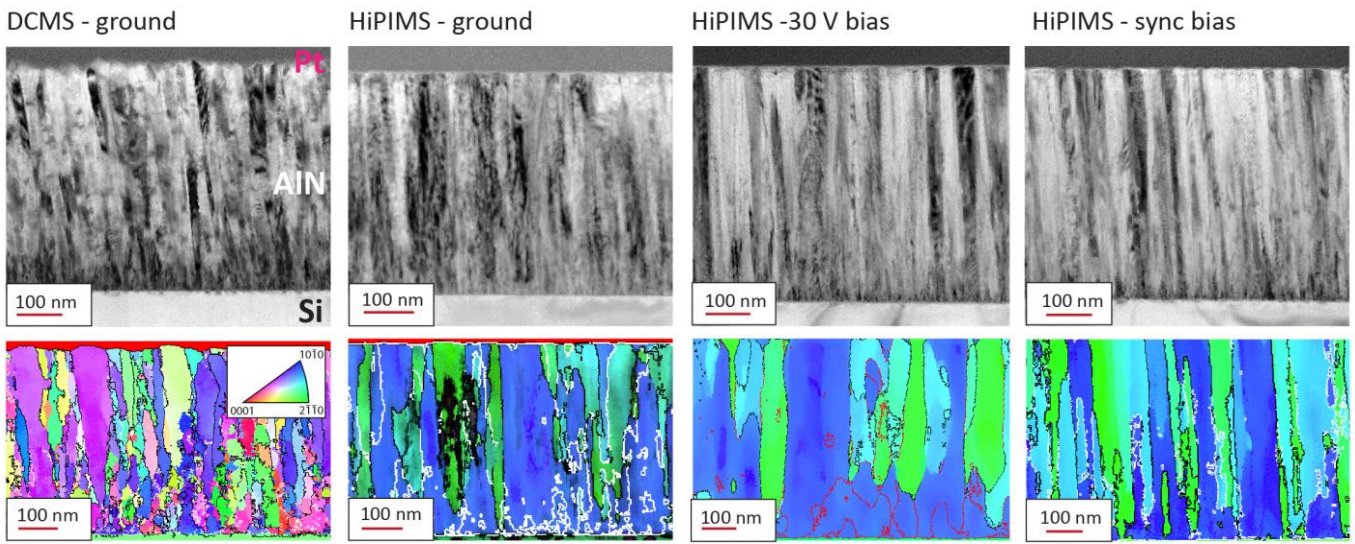

**(b) out-of-plane SPED imaging**

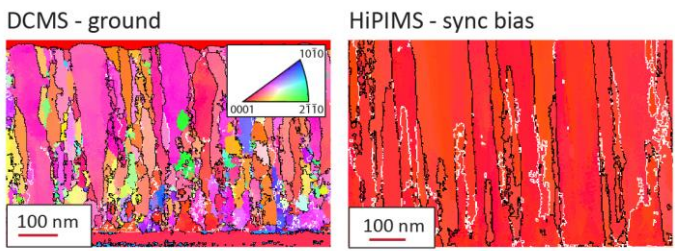

**(c) SAED**

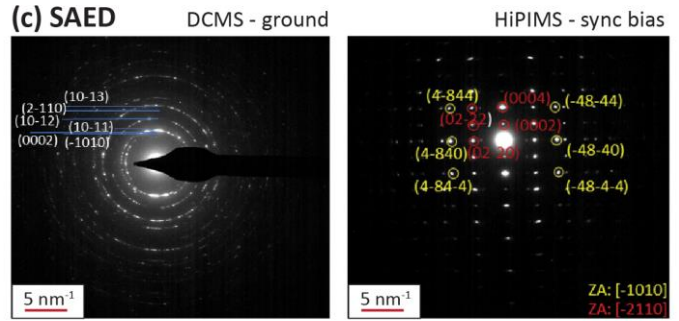

**Figure 8:** (a) Bright-field and SPED TEM imaging of four representative films. The red and white colors in the IPF map (inset of SPED images) represents low and high angle grain boundaries. The HiPIMS films show a pronounced columnar growth and out-of-plane orientation which are further improved with substrate biasing. A minor in-plane texture is evident for the HiPIMS deposited films. (b) Out-of-plane SPED imaging and (c) SAED patterns for DCMS and MIS-HiPIMS films. Both analyses show remarkable improvements in texture from DCMS to synchronized HiPIMS.

The difference in microstructure is also reflected in the surface roughness as evidenced by AFM measurements (see **Figure 9**). The root mean square (RMS) roughness measured over a 1 µm² area of the film is markedly reduced to less than 1 nm in HiPIMS-deposited films in contrast to the 4 nm RMS roughness for the DCMS film. The rough surface of the films deposited with grounded substrates in both, DCMS and HiPIMS, can be seen in the bright-field TEM cross-sections as well. Furthermore, to assess the polarization of the grains, qualitative PFM measurements were performed. Both the out-of-plane and the in-plane piezoresponse images revealed no variation in contrast in all tested films (see **Section S7 in supporting information**). The lack of apparent piezoelectric domains indicates a uniform polarization across all grains. These results are promising and indicate, that films deposited with HiPIMS, particularly with a synchronized substrate bias, could be used for piezoelectric applications, even when deposited via oblique angle deposition or on structured surfaces.

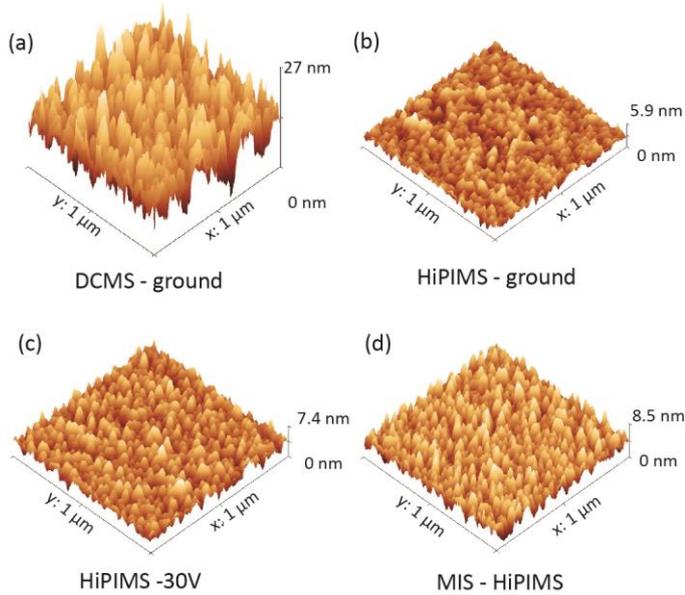

**Figure 9:** Surface topography of representative samples measured using AFM. The surface roughness is evidently lower in HiPIMS samples than the DCMS due to high ad atom mobility in case of HiPIMS.

### 3.5 Surface analysis and oxidation resistance

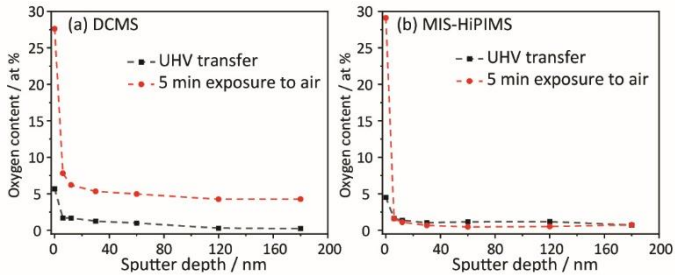

**Figure 10:** Depth profiling of DCMS and MIS-HiPIMS samples done in XPS for estimation of oxygen content in the films. In the case of DCMS deposition, the increased amount of oxygen in the deeper levels of films after exposure to air indicates grain boundary oxidation. Contrary to this, the MIS-HiPIMS film shows no oxidation due to its compact columnar structure.

Finally, the chemical composition of the films was analyzed via XPS combined with a UHV sample transfer, i.e. without exposing the films to the atmosphere. To probe the oxygen contamination in the films depth profiling was performed on the samples before and after exposure to atmosphere (see **Figure 10**). Both films show oxygen levels below 1 % in the film after a few minutes of sputtering. However, after performing the depth profiling on the same samples after exposure to the atmosphere, the DCMS film shows about ~ 5% of oxygen levels inside the films. Contrary to this, the HiPIMS films only show surface oxidation and stay oxygen-free in the bulk of the film. This is attributed to the compact microstructure of HiPIMS films, which prevents grain boundary oxidation of the films. The oxidation resistance offered by films deposited using synchronized HiPIMS makes them particularly interesting candidates for applications in reactive environments or with limited encapsulation. In addition, it was found based on the UHV XPS experiments that the films are stoichiometric within the margin of error of the measurement. The Al/N ratio was found to be $Al_{0.45}N_{0.55}$ on the slightly oxidized surface and $Al_{0.48}N_{0.52}$ after gentle Ar-ion milling (see supporting information, section S8). This composition was identical for DCMS and MIS-HiPIMS deposited films.

### 3.6 Static deposition and deposition on structured substrates– DCMS vs MIS-HiPIMS

Finally, to assess the influence of different growth modes during static deposition (i.e. without substrate rotation), the AlN films were deposited using DCMS with a grounded substrate and MIS-HiPIMS on four Si substrates placed at different positions on the substrate holder, corresponding to different incident deposition angles, as shown in **Figure 11(a-b)**. A closed magnetic configuration was used, to avoid influences from non-uniform plasma heating. The preferred growth axis of the films was analyzed using a χ-scan in XRD after fixing 2θ and ω on the AlN (0002) maximum. χ is referenced to the substrate normal, i.e. a deviation of peak position from 0° indicates a tilt of grains in the film.

The DCMS films give broader peaks as compared to the MIS-HiPIMS films indicating higher mosaicity and misalignment among the grains, in line with our expectation from the deposition using substrate rotation. Strikingly, the peak position shifts to a higher χ-angle in DCMS films as the deposition angle becomes higher. In contrast, the MIS-HiPIMS films' peaks nearly stay around 0° indicating the growth of grain is almost normal to the substrate plane. This confirms that the energetic ions in MIS-HiPIMS improves not only the texture of the growing film but can also be utilized to improve the out-of-plane orientation at higher sputter angles during static deposition (see Figure 11).

To further validate this, we deposited the films on a structured wafer of Si with substrate rotation "on" throughout the deposition to ensure uniform deposition. The etched (001) Si wafer exhibits pyramids with (111) facets (See **Figure 11(c)**). The χ-scan was performed (See **Figure 11(d)**) by aligning the XRD on the (111) reflection of Si and optimizing φ (i.e., the moving the (111) facets of pyramids into the diffraction plane). The deviation of the AlN peaks from the Si peak in the χ-scan effectively tells us the tilt angle of grains in the film with respect to the normal of the (111) facet. The MIS-HiPIMS films' grains shows a smaller misalignment from the substrate normal and sharper peak compared to the DCMS film. The result is in agreement with the previous investigations and is due to increased ad atom mobility and bombardment of ions normal to the substrate plane. The MIS-HiPIMS film still has a small amount of grains' tilt which is supposedly due to the combination of oblique-angle deposition and the structured surface. Nevertheless, based on these results, it can be concluded that MIS-HiPIMS is a powerful tool to obtain highly oriented and textured films irrespective of the deposition angle.

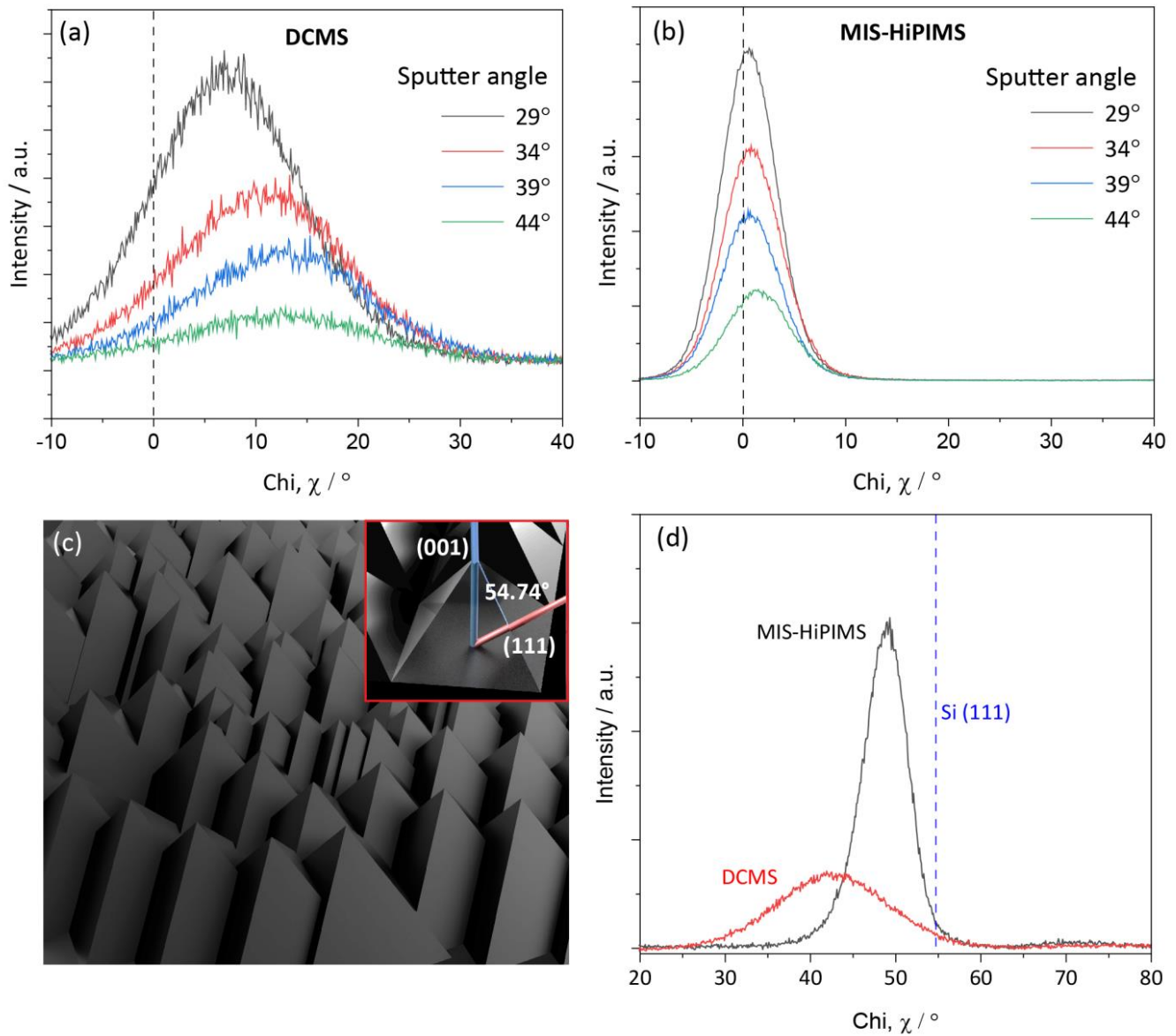

**Figure 11:** Deposition of AlN films on Si substrates placed at different sputter angles (without rotating the sample holder). χ-scan for **(a)** DCMS **(b)** MIS-HiPIMS - deposited films for different sputter angles. Higher deviation in peak positions in DCMS films indicates a higher tilt of grains. The peak further shifts to a larger angle as the deposition angle gets higher, while the MIS-HiPIMS films stay nearly around 0° and moreover at the same position. **(c)** Schematic of a structured Si substrate. The substrate has micro-pyramids etched over a (001) Si wafer and have facets oriented in (111) plane (zoomed in image shown inset). **(d)** χ-scan for DCMS and MIS-HiPIMS film deposited on structured Si substrate with substrate rotation on. The difference of χ angle between the Si peak and AlN peaks represents the tilt of grain with respect to the Si substrate normal, which is evidently higher for DCMS film.

# 4. Conclusions

In this study, the effect of different deposition approaches and techniques, namely, DCMS and HiPIMS, were discussed to grow textured AlN (0002) films in oblique angle deposition conditions. Specifically, we investigated if HiPIMS in combination with low negative substrate bias potentials can mitigate the detrimental effects of shallow deposition angles.

Plasma diagnostics and multiple process parameter optimizations were performed including a change of the magnetic configuration inside the deposition chamber as well as different substrate bias potentials. It was demonstrated that the plasma can be guided toward the substrate with an open-field magnetic configuration resulting in an increased ion-irradiation on the growing film. Energy and time-resolved QMS analysis was performed to characterize the kinetic energy of major ion species in the HiPIMS process. After measuring the time-of-flight of ions following the HiPIMS pulse, the substrate bias potential was pulsed and synchronized with the HiPIMS pulse to selectively accelerate the metal-ions on the growing film.

Four representative films grown with different approaches were chosen to compare the change in crystallinity and texture: DCMS with a grounded substrate, HiPIMS with a grounded, -30 V continuous, and a pulsed-synchronized substrate bias potential. We found that the crystalline quality and texture of the films significantly improved from DCMS to HiPIMS, owing to higher ad atom mobility due to energetic ion bombardment. The quality of films was further improved by increasing the kinetic energy of impinging ions with the help of substrate biasing. The films deposited with HiPIMS were found to have low surface roughness and an evident dense columnar growth in the (0002) direction, in contrast to the DCMS film. Overall the MIS-HiPIMS approach yielded the most promising results. Despite the low substrate bias potentials, some incorporation of Ar in the film was observed, resulting in high compressive stress. This effect however was reduced by applying the synchronized substrate bias potential. In addition to the improvements in crystallinity and texture, we found that a positive side effect of the compact microstructure is an improved oxidation resistance when the layers are exposed to ambient conditions.

The extent of the grains' tilt in the film was estimated with the help of χ-scans in XRD. The angle at which grains are aligned with respect to the substrate normal was found to be directly correlated with the angle of sputter flux in the DCMS process. Moreover, the columnar grains in the film stay nearly normal in MIS-HiPIMS film irrespective of the deposition angle. These results were further validated by deposition on a structured Si substrate.

Finally, it is concluded that synchronized HiPIMS can be a promising deposition approach to synthesize films with pronounced out-of-plane texture on non-uniform substrates or in oblique deposition conditions. While more work is needed to further reduce the Ar content and consequently the compressive stress in MIS-HiPIMS mode, the presented results already demonstrate that synchronized HiPIMS processes could open up exciting opportunities for the deposition of functional defect-sensitive thin film materials in the future.

# Acknowledgements

The authors would like to thank Ulrich Müller for his help during the setup of the deposition chamber. Help from Monalisa Ghosh during process characterization and Alexander Wieczorek for the image design of structured Si is gratefully acknowledged. J. P. acknowledges funding by the SNSF (project no. 200021_196980). S.Z. acknowledges funding by the Empa research commission. G.L. acknowledges the Swiss National Science Foundation (SNSF), project number 200021_192224 for financially supporting this research. M.T. acknowledges the financial support by the Swiss National Science Foundation under project No. 200021_188414. M.T. and M.F.S. acknowledge the Swiss National Science Foundation Spark funding CRSK-2_196061.

## Author contributions

**J. P.:** Conceptualization, Investigation, Formal analysis, Visualization, Writing - Original Draft; **A. S.:** Investigation, Formal Analysis, Writing – Review & Editing; **S. Z.:** Investigation, Writing – Review & Editing; **G.L.:** Formal Analysis, Writing – Review & Editing; **C.C.:** Formal Analysis, Writing – Review & Editing; **M.F.S.:** Investigation, Formal Analysis, Writing – Review & Editing; **M.T.:** Investigation, Formal Analysis, Writing – Review & Editing; **K.T.:** Investigation, Writing – Review & Editing; **J.M.:** Writing – Review & Editing; **S.S.:** Conceptualization, Supervision, Methodology, Formal analysis, Funding acquisition, Writing – Review & Editing

10.1063/1.126244.

**Supporting Information for**

# Improving the crystallinity and texture of oblique angle deposited AlN thin films using reactive synchronized HiPIMS

Jyotish Patidar[1], Amit Sharma[2], Siarhei Zhuk[1], Giacomo Lorenzin[3], Claudia Cancellieri[3], Martin F. Sarott[4], Morgan Trassin[4], Kerstin Thorwarth[1], Johann Michler[2], Sebastian Siol[1*]

[1]Laboratory for Surface Science and Coating Technologies, Empa – Swiss Federal Laboratories for Materials Science and Technology, Switzerland

[2]Laboratory for Mechanics of Materials and Nanostructures, Empa – Swiss Federal Laboratories for Materials Science and Technology, Switzerland

[3]Laboratory for Joining Technologies and Corrosion, Empa – Swiss Federal Laboratories for Materials Science and Technology, Switzerland

[4]Department of Materials, ETH Zürich, Switzerland

Corresponding author: sebastian.siol@empa.ch

Keywords:  HiPIMS, AlN, synchronized HiPIMS, texture, oblique-angle deposition

## S1: Schematic of gas diverter valves in the deposition system

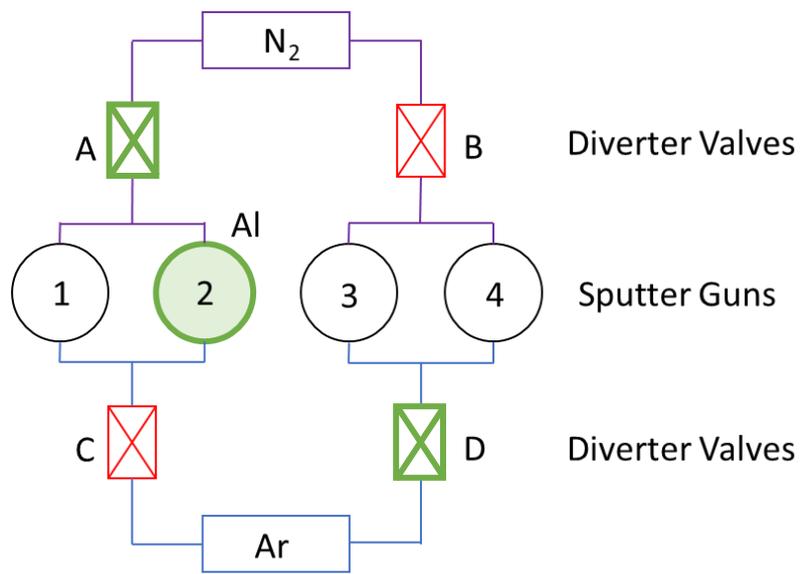

**Figure S1:** The deposition chamber has a unique set of diverter valves (DV) attached with the gas lines, which is used to route the process gas to specific sputter guns. For all the depositions in this study, A and D valves were kept open, i.e., the nitrogen gas was supplied directly to the Al target, while the argon gas was supplied away from it.

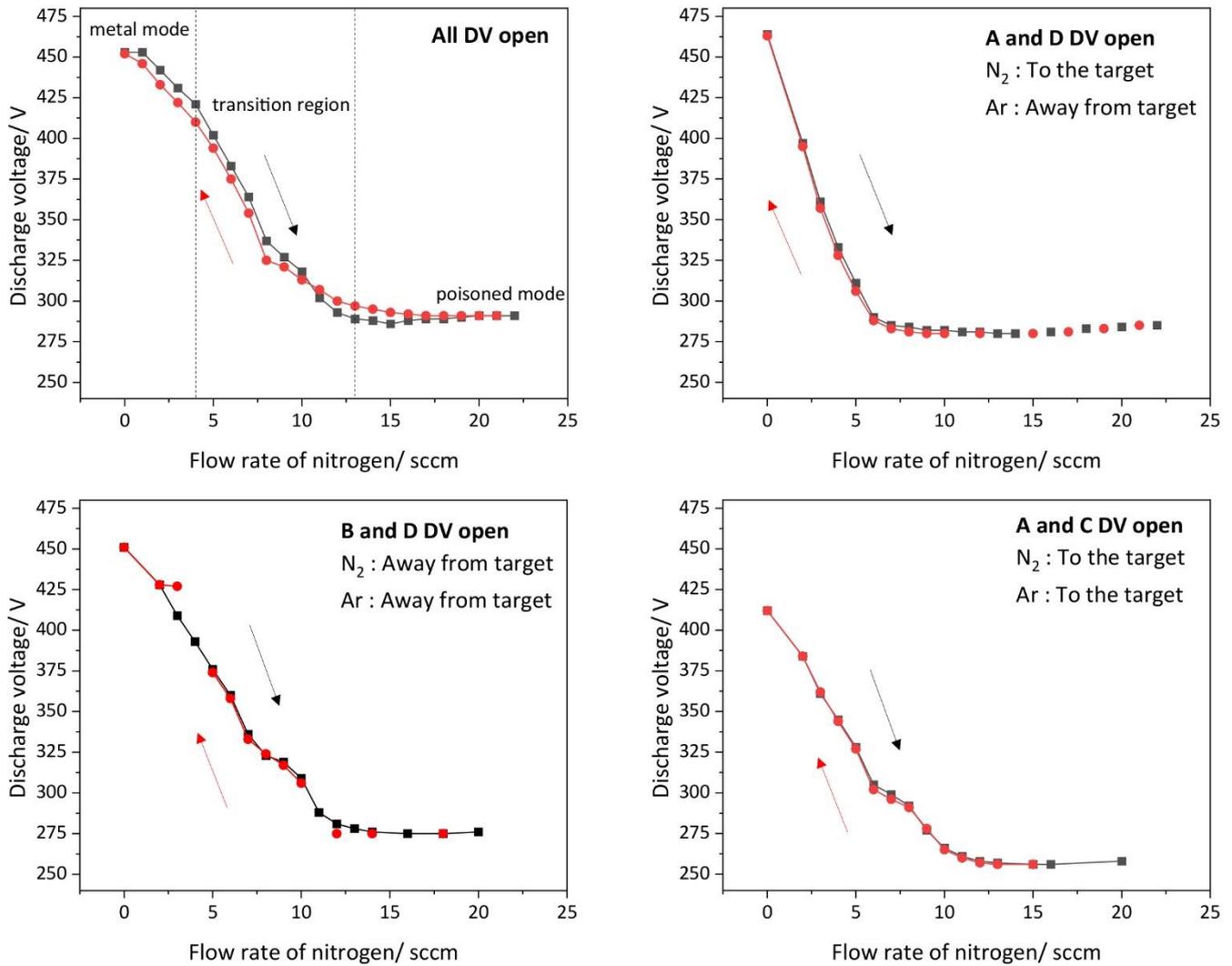

**Figure S2:** Hysteresis studies done using DCMS for Al target at 5 µbar, 100 W for different diverter valve (DV) settings: (a) Ar and $N_2$ supplied to all guns (b) Ar supplied away from the target and $N_2$ to the target (c) Both Ar and $N_2$ supplied away from the target (d) Both Ar and $N_2$ supplied to the target. Since a poisoned state of target was obtained at relatively lower flow rate of nitrogen with the (b) setup of valves, i.e. $N_2$ directed to the gun and Ar away from it, this condition was chosen for further investigations.

## S3: Time-resolved mass spectrometer measurements

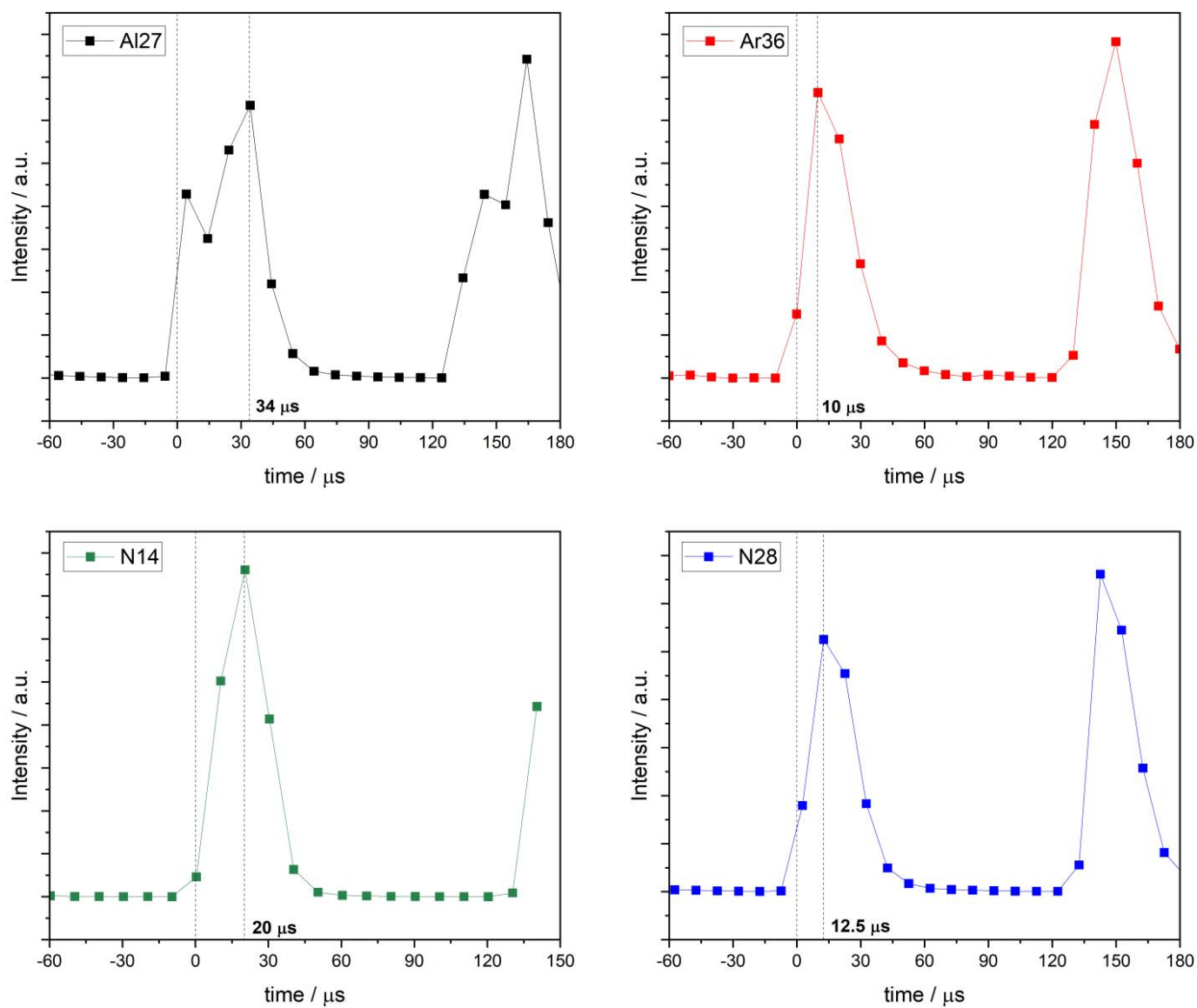

**Figure S3:** Time-resolved mass spectrometer measurement for (a) $^{27}Al^+$ (b) $^{36}Ar^+$ (c) $^{14}N^+$ (d) $^{28}N_2^+$ for a HiPIMS pulse. Here 0 is defined as the onset of HiPIMS pulse.

## S4: Rocking curves of closed-field-deposited samples

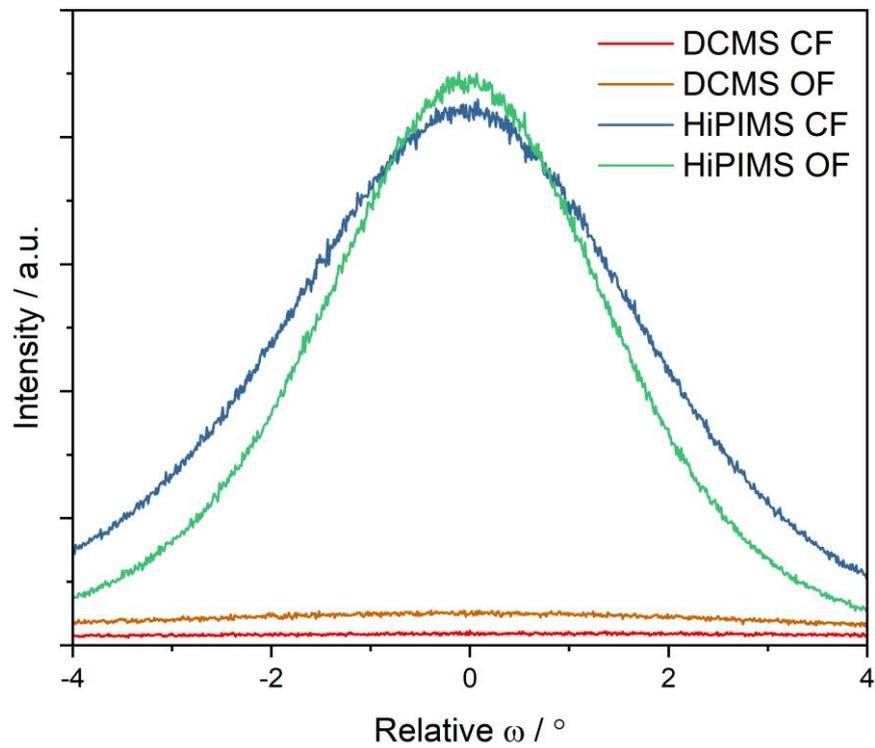

**Figure S4:** Rocking curve of samples deposited with open-field (OF) and closed-field (CF) magnetic configurations using DCMS and HiPIMS, all with grounded substrate.

## S5: Stress determination using crystallite group method (CGM)

The Crystallite Group Method (CGM) is a variation of the $\sin^2\psi$-method that takes into account the anisotropic mechanical response of the film to the stress.

Bragg's reflections used in this work to determine the stress are reported in the following table:

| Bragg's reflection | 2θ | ψ |
|---|---|---|
| (0002) | 36.10° | 0° |
| (10$\bar{1}$3) | 66.06° | 31.64° |
| (10$\bar{1}$2) | 49.83° | 42.74° |
| (10$\bar{1}$1) | 37.95° | 61.59° |
| (20$\bar{2}$3) | 94.88° | 50.94° |

The strain in the films for the calculation of stress is calculated using:

$$strain = \frac{d_{hkl} - d_{hkl}^o}{d_{hkl}^o},$$

where $d_{hkl}$ is the spacing of plane (hkl) and $d_{hkl}^o$ is the bulk nominal d value of (hkl) plane.

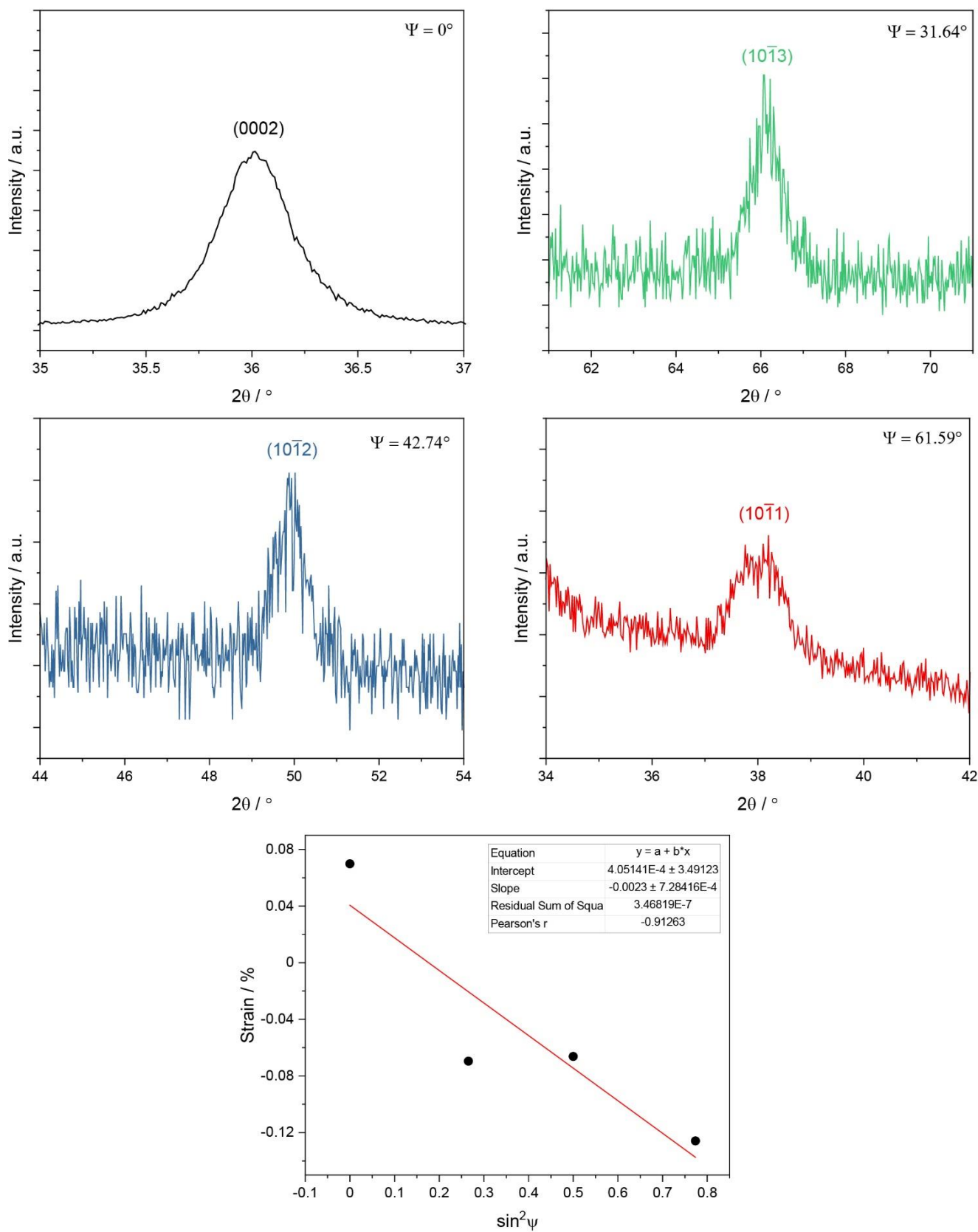

**Figure S5:** XRD patterns at different Ψ angles for AlN (0002) film deposited using DCMS with grounded substrate.

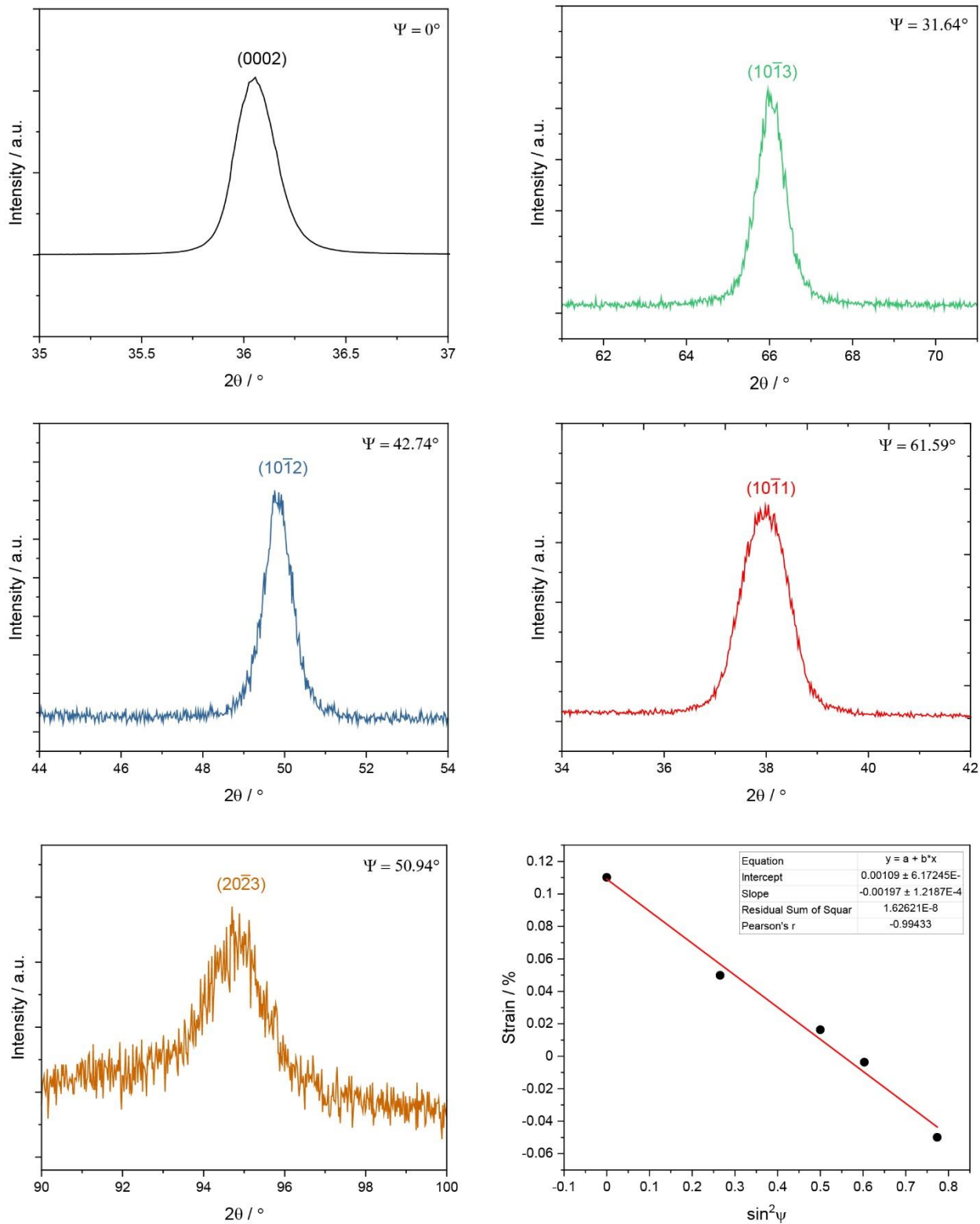

**Figure S6:** XRD patterns at different Ψ angles for AlN (0002) film deposited using HiPIMS with grounded substrate.

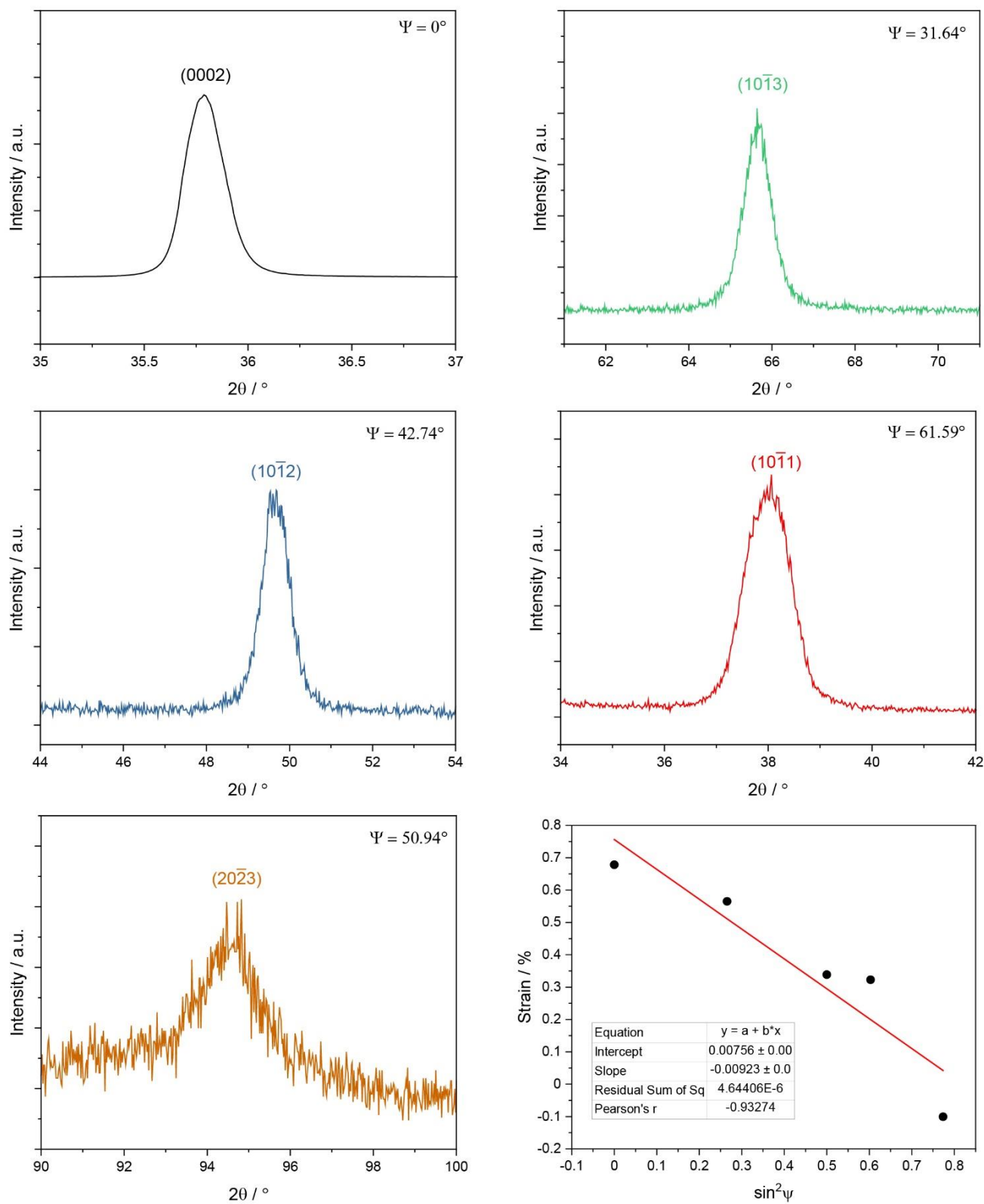

**Figure S7:** XRD patterns at different Ψ angles for AlN (0002) film deposited using HiPIMS with -30 V continuous DC substrate bias.

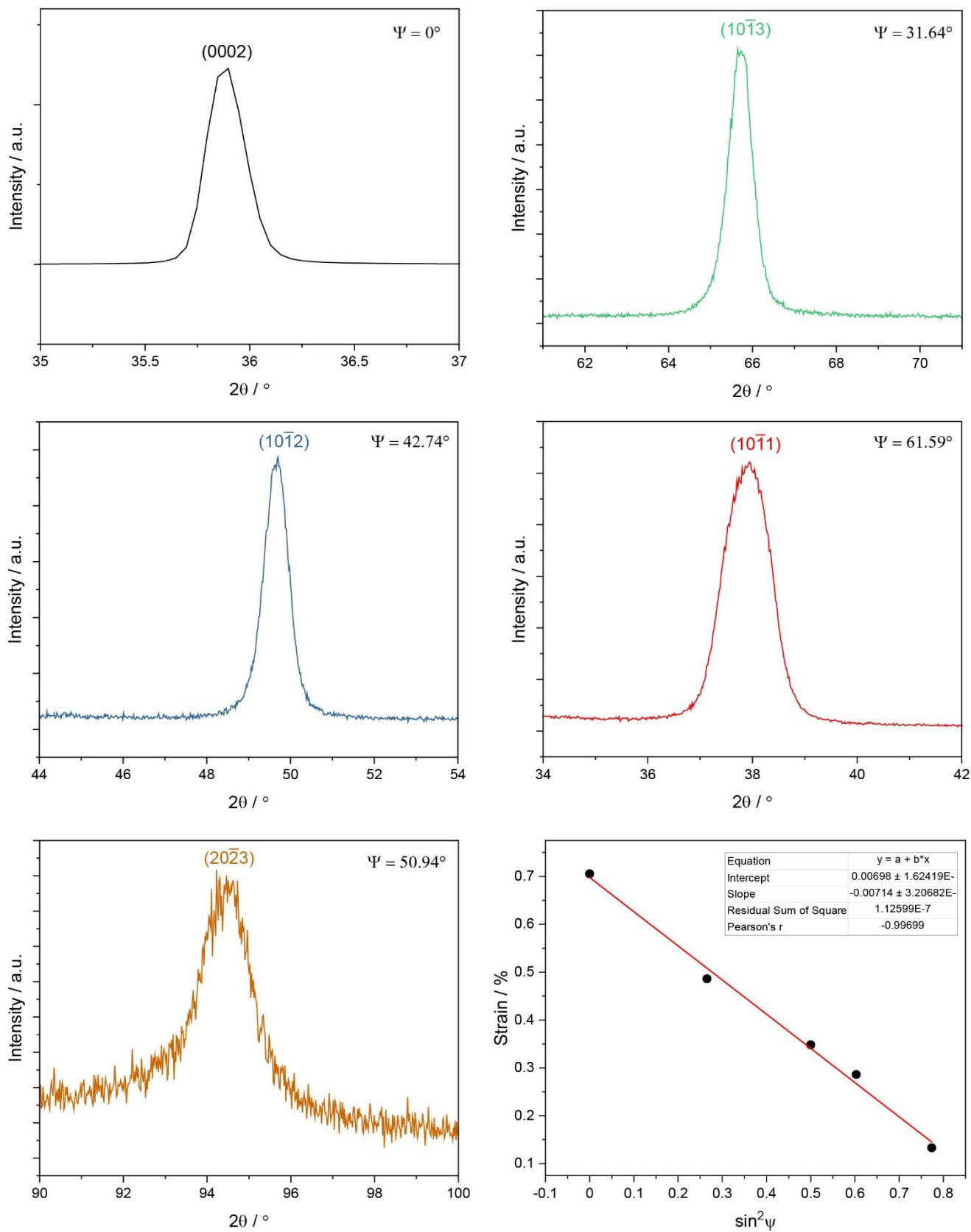

**Figure S8:** XRD patterns at different Ψ angles for AlN (0002) film deposited using HiPIMS with -30 V synchronized substrate pulse bias potential.

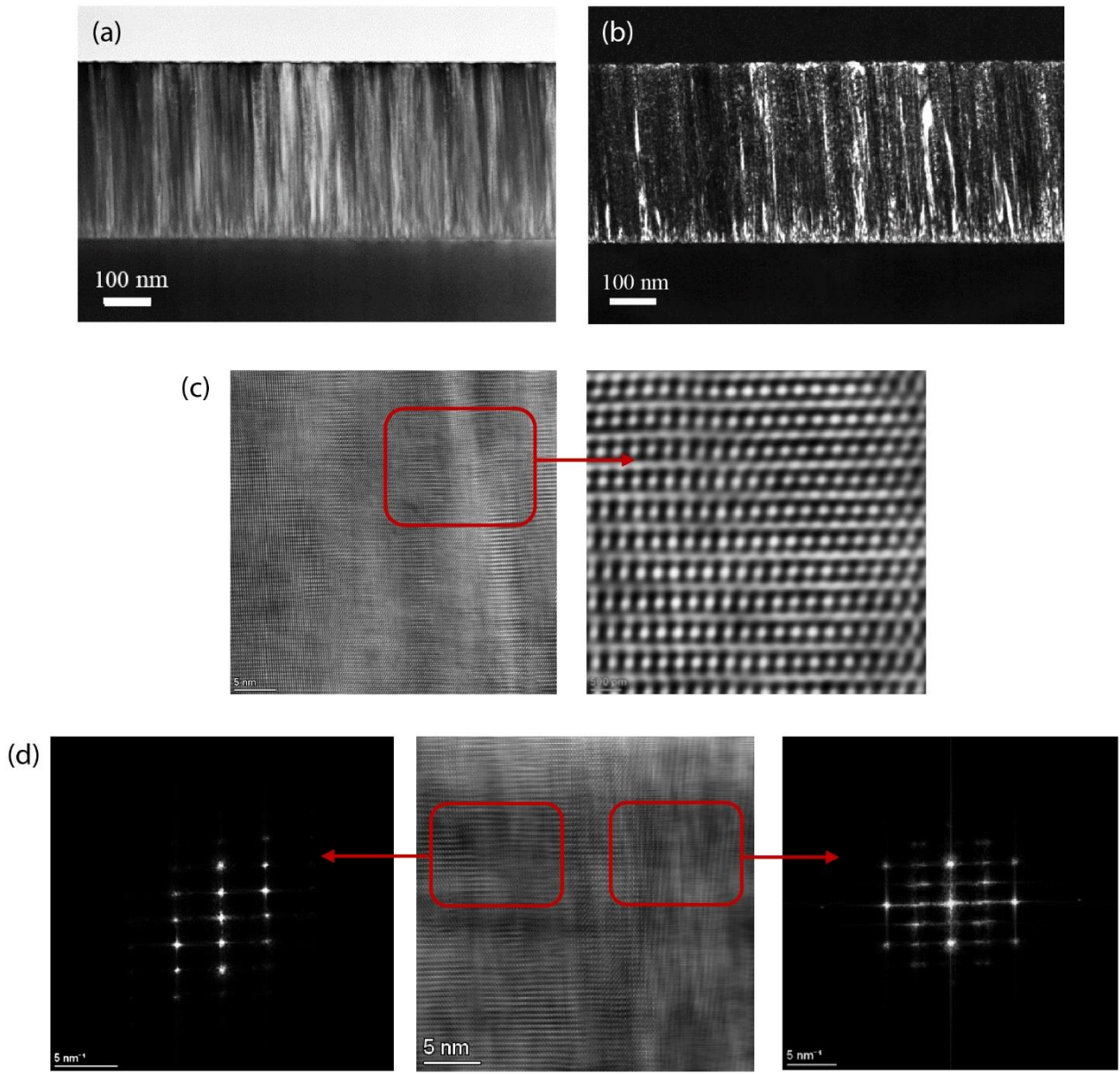

**Figure S9:** (a) High-angle annular dark field (HAADF) (b) TEM dark field (c) high-resolution TEM (HR-TEM) and (d) selected area electron diffraction (SAED) imaging of AlN (002) film deposited using MIS-HiPIMS.

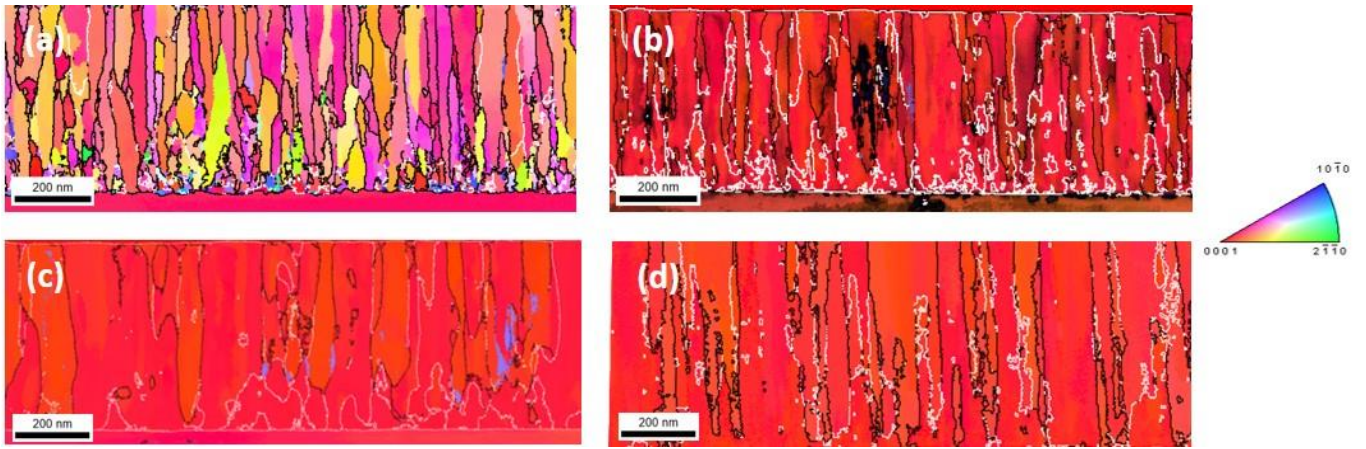

**Figure S10:** Out-of-plane SPED imaging of (a) DCMS with grounded substrate (b) HiPIMS with grounded substrate (c) HiPIMS with -30 V DC bias (d) HiPIMS with -30 V pulsed synchronized bias. All HiPIMS films show a pronounced (0002) orientation, while the DCMS film has grains oriented in different planes.

## Out-of-plane PFM

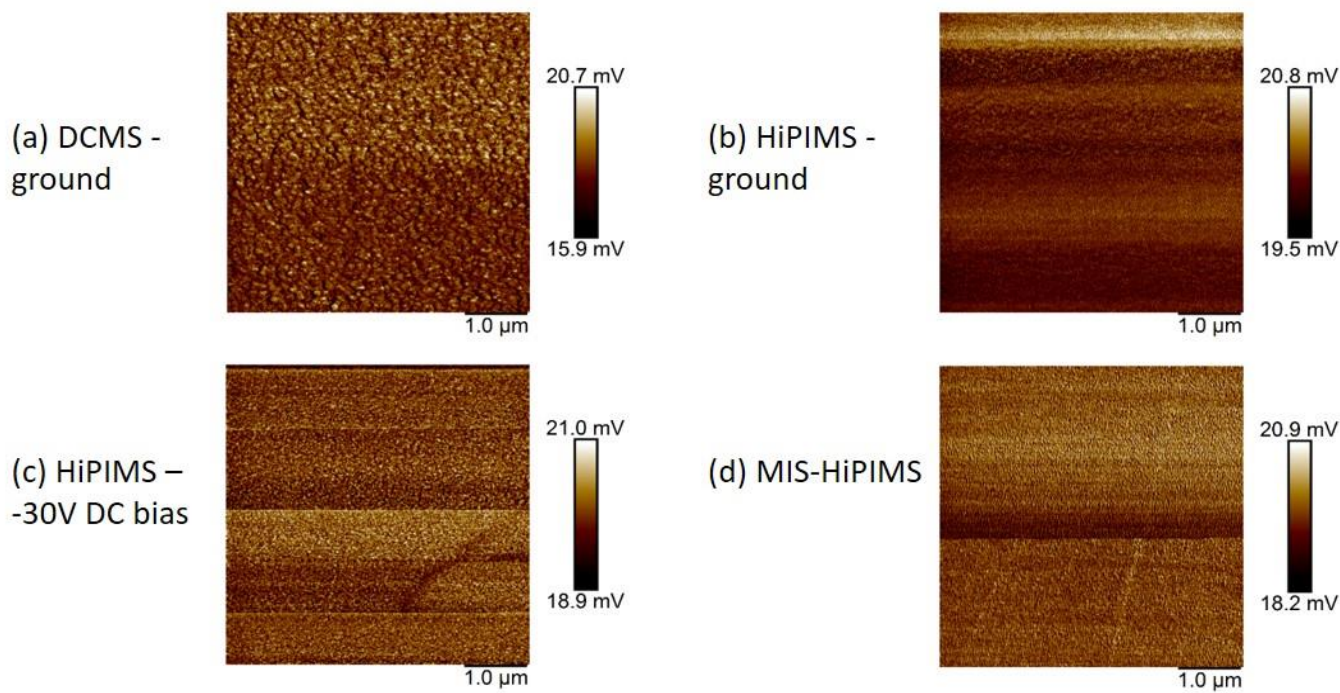

## In-plane PFM

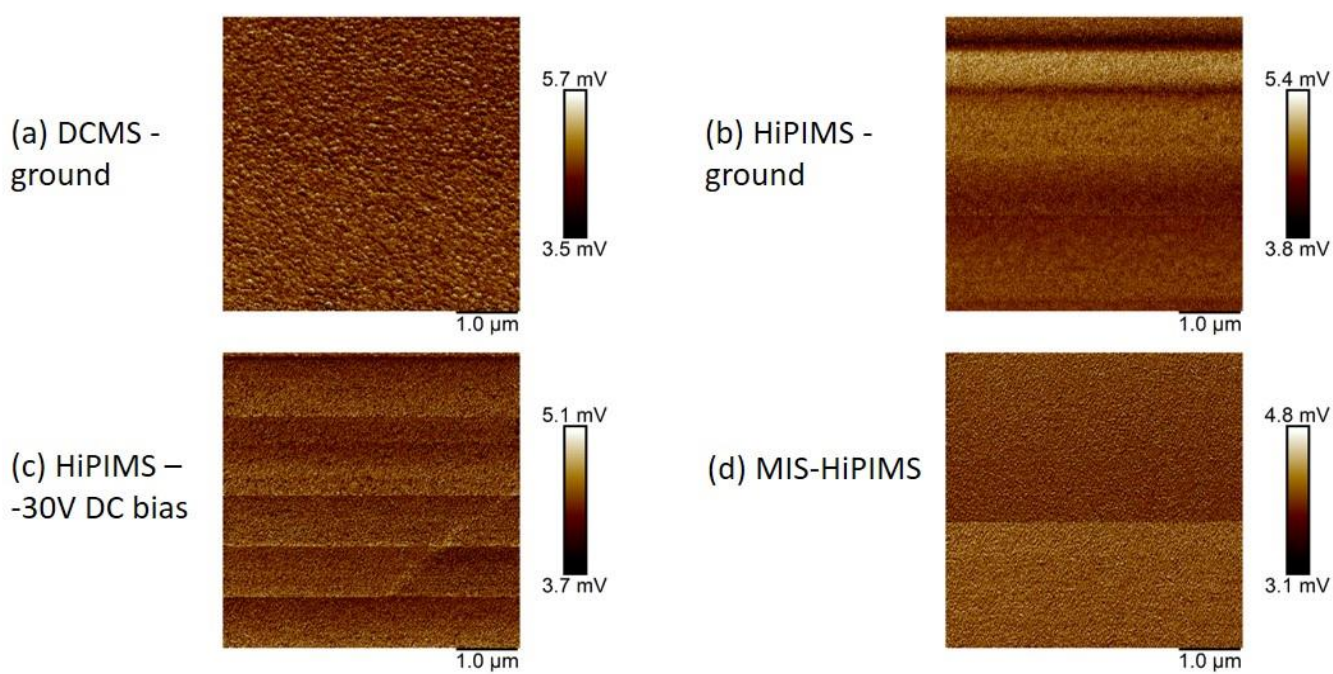

**Figure S11:** Piezoresponse force microscopy (PFM) of 4 representative samples discussed in this study. The films shows no contrast indicating uniform polarization in the films.

# S8: X-ray photoelectron spectroscopy (XPS)

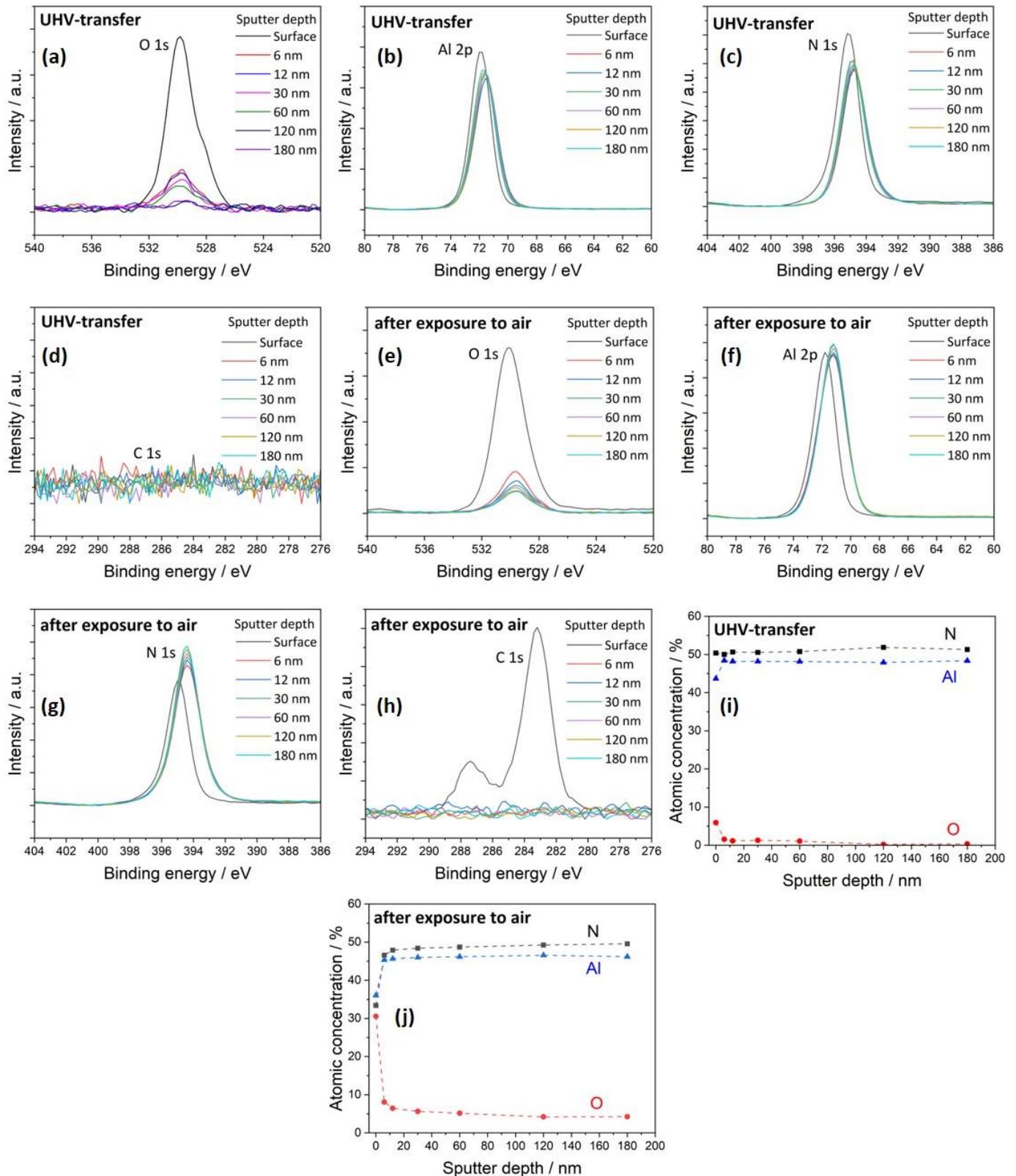

**Figure S12:** XPS depth profile analysis of UHV-transferred AlN films deposited using **DCMS** method for regions (a) O 1s (b) Al 2p (c) N 1s (d) C 1s and after exposing the films to air for regions (e) O 1s (f) Al 2p (g) N 1s (h) C 1s. The atomic concentration of N, Al and O are calculated based on the XPS spectra and is summarized in (i) for UHV-transfer and (j) after exposing to air with respect to the sputter depth.

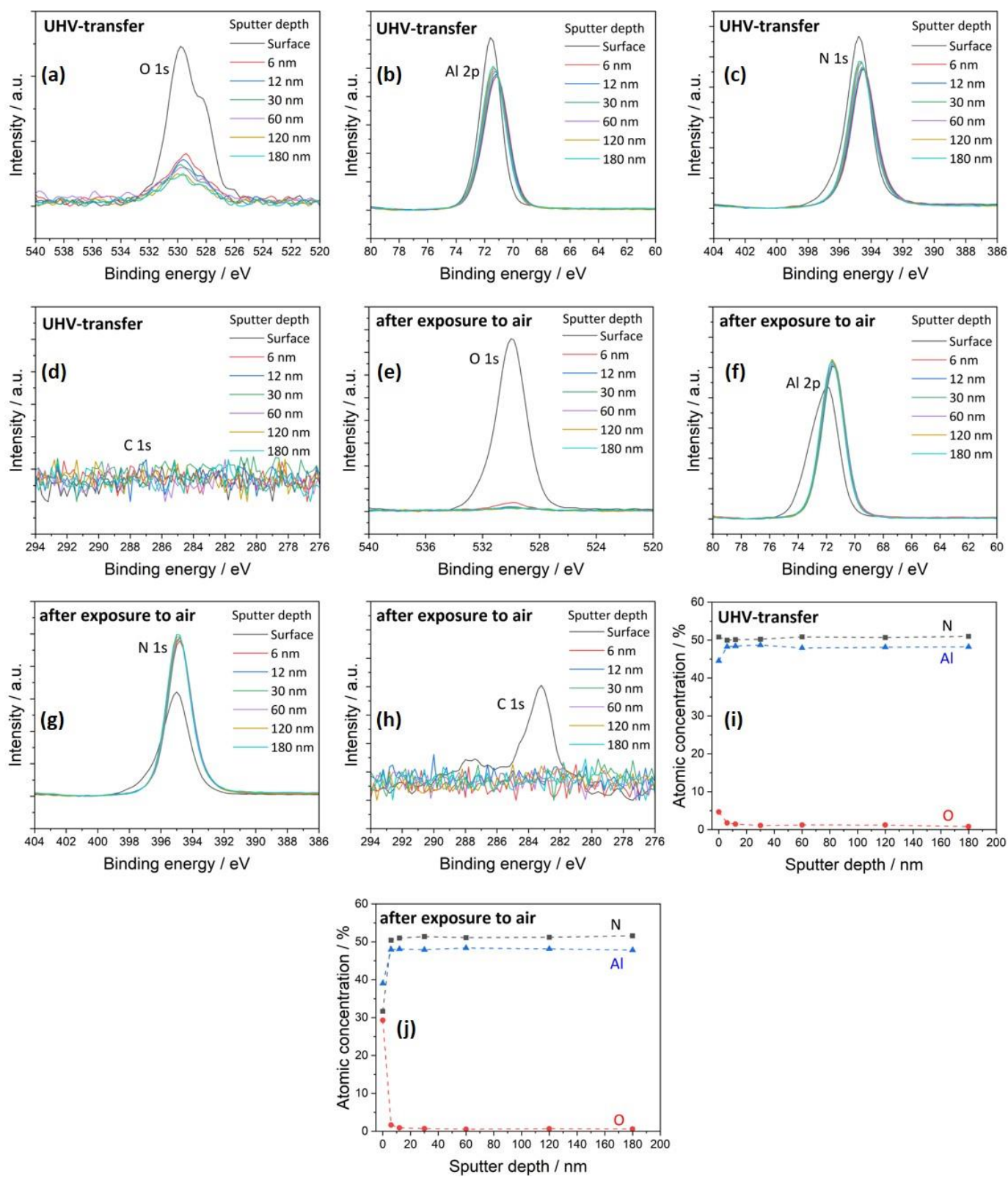

**Figure S13:** XPS depth profile analysis of UHV-transferred AlN films deposited using **MIS-HiPIMS** method for regions (a) O 1s (b) Al 2p (c) N 1s (d) C 1s and after exposing the films to air for regions (e) O 1s (f) Al 2p (g) N 1s (h) C 1s. The atomic concentration of N, Al and O are calculated based on the XPS spectra and is summarized in (i) for UHV-transfer and (j) after exposing to air with respect to the sputter depth.